\documentclass[a4paper,11pt]{scrartcl}

\usepackage[sort&compress,numbers]{natbib}
\usepackage{hyperref}
\usepackage{amssymb}
\usepackage{amsmath}
\usepackage{amsfonts}
\usepackage{graphicx}
\usepackage{dcolumn}
\usepackage{bm}
\usepackage{tikz}

\usetikzlibrary{decorations.pathmorphing}
\usetikzlibrary{math}
\begin{document}

\begin{titlepage}
\vspace{.3cm} \vspace{1cm}
\begin{center}
\baselineskip=16pt \centerline{\textbf{\Large{On Stability of}}}\vspace{.3truecm}
\centerline{\textbf{\Large{Asymptotically Free Mimetic Ho\v{r}ava Gravity}
}}
\vspace{1truecm}
\centerline{\large\textbf{  Tobias B. Russ$^{1} $}\ \ } \vspace{.5truecm}
\emph{\centerline{$^{1}$ Ludwig-Maxmillians-Universität, Theresienstr. 37, 80333 Munich, Germany }%
}\\
\emph{Contact: tobias.russ@physik.uni-muenchen.de}
\end{center}
\vspace{2cm}
\begin{center}
{\textbf{ Abstract}}
\end{center}

Asymptotically free mimetic gravity has been introduced as a proposal for a classical limiting curvature theory with the purpose of singularity resolution. It was found that in a spatially flat universe an initial stage of exponential expansion with graceful exit is a generic consequence, regardless of the matter content. 
In this work I will analyze linear stability of cosmological perturbations in such a model, considering only the degrees of freedom of pure mimetic gravity. I show that the addition of Ho\v{r}ava-gravity-like higher order spatial curvature terms can lift the gradient instability of scalar perturbations, even when the gradient term has the wrong sign throughout.
Calculating the primordial spectra of tensor and scalar perturbations in the simplest single component model, I find that the initially scale invariant spectra turn out to be destroyed later by the rapidly varying speed of sound at horizon exit.

\end{titlepage}%

% \preprint{APS/123-QED}

% \tableofcontents
%%%%%%%%%%%%%%%%%%%%%%%%%%%%%%%%%%%%%%%%%%%%%%%%%%%%%%%%%%%%%%%%%%%%%%
\section{Introduction}

By the unique status of general relativity, any alternative theory of metric gravity usually either has to allow for higher derivatives of the metric, higher dimensions of spacetime or to introduce new fields, separate from the metric. 
Another option is to reparametrize the degrees of freedom of the physical metric itself, e.g. by a disformal transformation \cite{Bekenstein:1992pj}. 
``Mimetic gravity'' stems from the reparametrization of the physical metric $g_{\mu\nu}$ in terms of an auxiliary metric $\tilde{g}_{\mu\nu}$ and the ``mimetic field'' $\phi$ as
\begin{equation}
g_{\mu\nu}=\tilde{g}_{\mu\nu}        \tilde{g}^{\alpha\beta}\phi_{,\alpha}\phi_{,\beta}. \label{5:eq:mimetic_disf}
\end{equation}
This particular disformal transformation is special for two reasons: \textit{1.)} It is singular, explaining how a simple reparametrization can actually lead to new physics  \cite{Deruelle:2014zza}, \cite{Domenech:2015tca}, \textit{2.)} The physical metric is invariant under Weyl transformations of the auxiliary metric. This means that the new degree of freedom introduced by $\phi$ represents what was called a ``conformal degree of freedom of gravity''. 
Soon after this reparametrization was first introduced in \cite{Chamseddine:2013kea},
it was shown in \cite{Golovnev:2013jxa} that the mimetic field can be introduced equivalently as a constrained scalar field, subject to the constraint
\begin{equation}
g^{\mu\nu}\phi_{,\mu}\phi_{,\nu}=1.\label{5:eq:mimetic_constraint}
\end{equation}

Apart from the dust-like component called ``Mimetic Dark Matter'' that emerges as a constant of integration in the modified Einstein equation of mimetic gravity, the introduction of the mimetic field also enables a wealth of possible new terms in the gravity action.
By breaking shift symmetry in $\phi$, i.e. introducing a $\phi$ dependent potential, one can produce an extremely flexible theory where essentially any conceivable background solution can be realized in a Friedmann universe, cf. \cite{Chamseddine:2014vna}.

Conversely, if we restrict to shift symmetric theories without higher derivatives of the metric in the modified Einstein equation, the range of possibilities for a non-singular universe becomes very narrow. 
% Conversely, in shift symmetric theories without higher derivatives of the metric in the modified Einstein equation the possibilities for a non-singular modified background solution are very restricted. 
For the most natural (and arguably only viable) class of modified flat Friedmann universes, it was shown in \cite{Chamseddine:2019bcn}, \cite{Chamseddine:2019fog} that the only thing that can replace the Big Bang singularity is a smooth transition to a piece of de Sitter spacetime at limiting curvature. In this work I will show that this class of modifications also happens to coincide with the class of models that avoid a ghost instability of scalar metric perturbations.

The mimetic field, by definition (\ref{5:eq:mimetic_constraint}), provides a global time function whose gradient is everywhere timelike. In \cite{Chamseddine:2019gjh} it was shown that this can be used to covariantly dissect any scalar quantity that is invariant under spatial diffeomorphisms in the slicing given by $\phi$.
In this way it is easy to write down a Ho\v{r}ava-gravity-like theory with only higher spatial derivatives but no higher time derivatives or mixed derivatives. 
In Ho\v{r}ava gravity \cite{Horava:2009uw} such an ``asymmetry'' between space and time is used to improve the UV behaviour of the graviton propagator for the purpose of renormalizability. 
Projectable Ho\v{r}ava models have been shown to be renormalizable in \cite{Barvinsky:2015kil}, \cite{Barvinsky:2019rwn}.
Compared to other covariantized version of Ho\v{r}ava gravity like \cite{Germani:2009yt},  mimetic Ho\v{r}ava gravity has the advantage of not having any additional propagating degrees of freedom in a Minkowski background.
Interestingly, also the following connections between mimetic gravity and Ho\v{r}ava gravity can be drawn: In \cite{Mukohyama:2009mz} it was explored how a dust-like component emerges as a constant of integration in Ho\v{r}ava-Lifshitz gravity. In \cite{Ramazanov:2016xhp} an equivalence between the IR limit of projectable Ho\v{r}ava gravity and a mimetic matter scenario has been found. Another Ho\v{r}ava-like mimetic model has been presented in \cite{Cognola:2016gjy}.

In this paper I will show that higher spatial derivative terms of sixths order can not only render a power counting renormalizable theory, they also serve to alleviate the gradient instability of the scalar degree of freedom of mimetic gravity in an expanding universe.

The paper is organised as follows: 
In section \ref{5:sec:theory}, I merge parts of the Lagrangians from \cite{Chamseddine:2019fog} and \cite{Chamseddine:2019gjh} to introduce the theory that will be used in the rest of the paper. 
In section \ref{5:sec:background}, I re-derive a simplified version of the background solutions found in \cite{Chamseddine:2019bcn}. Introducing modified conformal time, these solutions can be written in closed form.
In section \ref{5:sec:perturbations}, I analyse metric perturbations in a flat Friedmann universe in comoving gauge. I discuss stability issues and calculate the primordial spectra of tensor and scalar perturbations for the particular case of a radiation dominated background. The analysis of the Mukhanov-Sasaki equations with modified dispersion relations follows similar steps as \cite{Bianco:2016yib}.
In section \ref{5:sec:conclusions}, I summarize my results and give a brief outlook on possible extensions. 
In Appendix \ref{5:sec:appA}, I present the second order actions for a more general mimetic theory and for the more general case of perturbations around a non-flat Friedmann universe. 
In Appendix \ref{5:sec:appB}, I perform the linear stability analysis for bouncing solutions driven by higher order spatial curvature terms in a non-flat universe, as found in \cite{Chamseddine:2019fog}.
Throughout this paper I use Planck units where  $G_0= G(\Box\phi=0)=1$, $\hbar=1$, $c=1$, $k_B=1$.
%%%%%%%%%%%%%%%%%%%%%%%%%%%%%%%%%%%%%%%%%%%%%%%%%%%%%%%%%%%%%%
\section{The theory}
\label{5:sec:theory}
Consider the shift-symmetric theory of mimetic gravity defined by
\begin{equation}
\mathcal{S}_{g}=-\frac{1}{16\pi}\int\textup{d}^{4}x\sqrt{-g}\left\{\mathcal{L}_{\textup{nhd}} + \mathcal{L}_{\textup{hd}}+\lambda\left(  g^{\mu\nu}\phi_{,\mu}%
\phi_{,\nu}-1\right) \right\},   \label{5:eq:S}%
\end{equation}
where the Lagrangian $ \mathcal{L} = \mathcal{L}_{\textup{nhd}} + \mathcal{L}_{\textup{hd}}$
is divided into a part $\mathcal{L}_{\textup{nhd}}$ without higher derivatives and a part $\mathcal{L}_{\textup{hd}}$ with higher spatial derivatives in the corresponding second order actions.

In \cite{Chamseddine:2019fog} we found that the Lagrangian
\begin{equation}
    \mathcal{L}_{\textup{nhd}} := f(\Box\phi)R+ (f(\Box\phi)-1) \widetilde{R}  +2\Lambda(\Box\phi)
    \label{5:eq:Lnhd}
\end{equation}
with
\begin{equation*}
\widetilde{R}:=  2\phi^{,\mu}\phi^{,\nu}G_{\mu\nu} - (\Box \phi)^2 + \nabla^{\mu}\nabla^{\nu}\phi \nabla_{\mu}\nabla_{\nu}\phi,
\end{equation*}
leads to a modified Einstein equation that is free of all higher derivatives of the metric.
Using the mimetic constraint, a slight generalization of a calculation from \cite{Chamseddine:2019gjh} shows that (apart from considerations involving boundary terms) it is equivalent to consider the Lagrangian
\begin{align}
    \mathcal{L}_{\textup{nhd}} &\doteq \tfrac{4}{3}\ell(\Box \phi)-f(\Box\phi)\left(\nabla^{\mu}\nabla^{\nu} \phi\nabla_{\mu}\nabla_{\nu}\phi -\tfrac{1}{3} (\Box \phi)^2 \right) -\widetilde{R}, \label{5:eq:Lnhd2}
\end{align}
where $\doteq$ denotes equality up to a total covariant derivative and
\begin{equation}
    \ell(\Box \phi) = \tfrac{3}{2}\Box \phi F(\Box \phi) - (\Box \phi)^2 f(\Box \phi) + \tfrac{3}{2}\Lambda(\Box \phi). 
\end{equation}
The function $F(\Box \phi)$ is definded by $F'(\Box \phi)\equiv \mathrm{d}F/\mathrm{d} \Box \phi  =f(\Box \phi)$.
In the following, the choice
\begin{equation}
    \Lambda(\Box \phi) =\tfrac{2}{3}\,\Box \phi \left (\Box \phi f(\Box \phi)\,-F(\Box \phi) \right ) 
    \label{5:eq:simplifying}
\end{equation}
will prove to be particularly simple.

Motivated by the goal of renormalizability along the lines of Ho\v{r}ava gravity \cite{Horava:2009uw}, as suggested in \cite{Chamseddine:2019gjh}, I include the sixth order higher spatial derivative terms
\begin{equation}
    \mathcal{L}_{\textup{hd}} := \sigma_T^4\, \widetilde{C}^{\mu}_{\nu}\widetilde{C}^{\nu}_{\mu} -\frac{\sigma_S^4}{8} P^{\mu}_{\nu}\nabla_\mu\widetilde{R}\nabla^{\nu}\widetilde{R},
    \label{5:eq:Lhd}
\end{equation}
where the projector $P_{\mu}^{\nu}=\delta_{\mu}^{\nu}-\phi_{,\mu}\phi^{,\nu}$ and the covariant analogues of the spatial Cotton tensor $\widetilde{C}^{\mu}_{\nu}$ and the spatial Ricci tensor $\widetilde{R}_{\mu\nu}$ have been introduced in \cite{Chamseddine:2019gjh} as
\begin{align}
\widetilde{C}_{\nu}^{\mu}&:=-\tfrac{1}{\sqrt{-g}}\epsilon^{\mu\rho\kappa\lambda
}\nabla_{\lambda}\phi\, \nabla_{\rho}\left(  \widetilde{R}_{\nu\kappa
}-\frac{1}{4}g_{\nu\kappa}\widetilde{R}\right), 
\label{5:eq:Cotton} \\
\widetilde{R}_{\mu\nu} &:=P_{\mu}^{\alpha}P_{\nu}^{\beta}R_{\alpha\beta}+\nabla_{\alpha}\left(\phi^{,\alpha}\nabla_{\mu}\nabla_{\nu}\phi\right).
\end{align}
Note that $g^{\mu\nu}\widetilde{R}_{\mu\nu} = - \widetilde{R}$,  using the mimetic constraint.

\paragraph{NB.} In a flat Friedmann universe, the Lagrangian studied in \cite{Chamseddine:2019bcn},
\begin{equation}
\begin{aligned}
  \widetilde{\mathcal{L}} &=  f(\Box\phi) R +  2\Lambda(\Box\phi) \\
  &\doteq \tfrac{4}{3}\ell(\Box \phi)-f(\Box\phi)\left(\nabla^{\mu}\nabla^{\nu} \phi\nabla_{\mu}\nabla_{\nu}\phi -\tfrac{1}{3} (\Box \phi)^2+\widetilde{R}\right),
\end{aligned}
\label{5:eq:Lhmd}
\end{equation}
leads to the same background dynamics as (\ref{5:eq:S}). However, in general its modified Einstein equation contains higher mixed derivatives of the metric in the synchronous frame $t=\phi$.
While it was found in \cite{Zheng:2017qfs} that these mixed derivatives can change the sign of the gradient term of scalar metric perturbations and help to prevent a gradient instability, it was later realized in \cite{Zheng:2018cuc}, \cite{Ganz:2019vre} that this comes at the price of introducing an additional  hidden degree of freedom. Even tough this second scalar degree of freedom does not show up when perturbing around a homogeneous background in unitary gauge, it was found in \cite{Ganz:2019vre} that already for perturbations around Minkowski spacetime with non-homogeneous mimetic field profile it can lead to instabilities. 
Since this type of higher mixed derivatives is not present in $\mathcal{L}_{\textup{nhd}}$, one could hope that no such additional degree of freedom will appear in this theory. A full Hamiltonian analysis similar to \cite{Zheng:2018cuc}, \cite{Ganz:2019vre}, \cite{Takahashi:2017pje}, \cite{Ganz:2018mqi} would of course be a more involved task beyond the scope of this paper.

%%%%%%%%%%%%%%%%%%%%%%%%%%%%%%%%%%%%%%%%%%%%%%%%%%%%%%%%%%%%%%%%%%%%%%

%%%%%%%%%%%%%%%%%%%%%%%%%%%%%%%%%%%%%%%%%%%%%%%%%%%%%%%%%%%%%%%%%%%%%%
\clearpage
\section{Background dynamics}
\label{5:sec:background}
%%%%%%%%%%%%%%%%%%%%%%%%%%%%%%%%%%%%%%%%%%%%%%%%%%%%%%%%%%%%%%%%%%%%%%
In the homogeneous, isotropic background given by the flat Friedmann metric
\begin{equation}
    \mathrm{d}s^2 = dt^2-a^2(t) \delta_{ij}\mathrm{d}x^i \mathrm{d}x^j, 
    \label{5:eq:flatFriedmann}
\end{equation}
the only consistent background solution of the mimetic constraint up to shifts is
\begin{equation}
    \phi = t,
    \label{5:eq:backgroundphi}
\end{equation}
for which we find
\begin{equation}
    \Box \phi \equiv \kappa = 3 H \equiv 3 \tfrac{\dot{a}}{a}.
\end{equation}
Either from the equation of motion given in \cite{Chamseddine:2019fog} or from analysis of the zeroth order action (see appendix \ref{5:sec:appA}), we arrive at the modified Friedmann equation
\begin{equation}
 \tfrac{2}{3}\left(\kappa\ell'(\kappa)-\ell(\kappa)\right) =\frac{c_{\textup{MDM}}}{a^3}+8\pi  \varepsilon^{m} =: 8\pi \varepsilon.
 \label{5:eq:mFriedmann}
\end{equation}
The constant of integration $c_{\textup{MDM}}$ describes the contribution of mimetic matter and $\varepsilon^m$ is some general homogeneous, isotropic matter energy density. Note that this background equation is the same for all the theories (\ref{5:eq:S}), (\ref{5:eq:Lnhd2}) and (\ref{5:eq:Lhmd}).
Using the simplifying choice (\ref{5:eq:simplifying}) and the suggestive notation $G = 1/f$ familiar from \cite{Chamseddine:2019bcn}, the modified Friedmann equation becomes
\begin{equation}
  H^2 =  \frac{8\pi}{3} G(\kappa)\, \varepsilon.
  \label{5:eq:simiplifiedMF}
\end{equation}
Assuming a monotonically decreasing dependence of $\varepsilon$ on the scale factor, such a modified Friedmann equation can be understood as an integral curve of the form $a (H^2)$ in the phase space spanned by $a$ and $H$, cf. \cite{Chamseddine:2019fog}, \cite{deCesare:2019pqj}.
The only possible relations of this form which \textit{a)} are one-to-one
\textit{b)} have limiting curvature $\kappa<\kappa_0$
and \textit{c)} obey the GR-limit $G(\kappa) = 1+\mathcal{O}\left((\kappa/\kappa_0)^2\right)$, 
generically replace the Big Bang singularity by a smooth transition to an initial de Sitter stage. In (\ref{5:eq:simiplifiedMF}) such a behaviour can only be realized by ``asymptotic freedom''\footnote{Note that in (\ref{5:eq:mFriedmann}) without using (\ref{5:eq:simplifying}), such a background behaviour could be equally well implemented by a choice of $\Lambda$ without asymptotic freedom. However, as shown in \cite{Chamseddine:2019bcn}, \cite{Chamseddine:2019fog}, asymptotic freedom becomes unavoidable for singularity resolution in an anisotropic universe.}, i.e. $G(\kappa \to \kappa_0) \to 0$. 
For a concrete example with limiting curvature $\kappa_0$, I will take the simple choice
\begin{equation}
   G(\kappa) =  f(\kappa)^{-1} = 1-\left(\frac{\kappa}{\kappa_0}\right)^2,
    \label{5:eq:f}
\end{equation}
familiar from \cite{Chamseddine:2019bcn}. Note that in this case $\ell'' = f^2$. Assuming a single matter component with constant equation of state $w=p/\varepsilon$, taking a time derivative of (\ref{5:eq:simiplifiedMF}) we find that during the inflationary stage
\begin{equation}
     -\frac{\dot{H}}{H^2}  = \frac{3(1+w)}{2} G(\kappa)\ll 1,
\end{equation}
where $\dot{}$ denotes $t$ derivatives.
However, the second ``slow-roll'' parameter
\begin{equation}
    \frac{\ddot{H}}{2H\dot{H}} = \frac{3(1+w)}{2} = \mathcal{O}(1)
\end{equation}
is constant and of order of unity, showing that this background solution does not fit into a ``slow-roll'' description.

It is straightforward to obtain the following implicit solution of (\ref{5:eq:simiplifiedMF}) for $\kappa(t)$:
\begin{equation}
    \frac{1+w}{2}\kappa_{0}\,t = \frac{\kappa_0}{\kappa} - \mathrm{atanh}\frac{\kappa}{\kappa_0}
\end{equation}

%%%%%%%%%%%%%%%%%%%%%%%%%%%%%%%%%%%%%%%%%%%%%%%%%%%%%%%%%%%%%%%%%%%%%%%%%%%%
\paragraph{Modified conformal time.}
Introducing the modified conformal time coordinate $\tilde{\eta}$ by
\begin{equation}
    \mathrm{d} \tilde{\eta} = \frac{\mathrm{d}t}{a\sqrt{f}},
\end{equation}
the modified Friedmann equation (\ref{5:eq:simiplifiedMF}) in modified conformal time $\tilde{\eta}$ looks exactly like the usual Friedmann equation in usual conformal time:
\begin{equation}
    \left(a_{\tilde{\eta}}\right)^2 = \frac{8\pi}{3} \varepsilon \,a^4 
\end{equation}
A subscript $a_{\tilde{\eta}} = \partial_{\tilde{\eta}}a$ denotes $\tilde{\eta}$ derivatives. 
Assume that the total energy density is dominated by a component with equation of state $w$ and parametrized as 
\begin{equation}
    \frac{8\pi}{3} \varepsilon = \left(\frac{c}{H_l^2}\right)^{\tfrac{1+3w}{2}} \left(\frac{1+3w}{2} \,a\right)^{-3(1+w)},
    \label{5:eq:etot}
\end{equation}
where the prefactors were introduced for later convenience and $H_l = \kappa_0/3$ accounts for dimensions such that $c$ is dimensionless. In this case we find the solution
\begin{equation}
    a(\tilde{\eta}) = \frac{2}{(1+3w)} \frac{\sqrt{c}}{H_l} \,\tilde{\eta}^{\frac{2}{1+3w}},
    \label{5:eq:mfsolution}
\end{equation}
where the initial condition $a(\tilde{\eta}=0)=0$ was used. The range of modified conformal time is then $0<\tilde{\eta}<\infty$.

For the choice (\ref{5:eq:f}) the solution for $\kappa(\tilde{\eta})$ is
\begin{equation}
    \left(\frac{\kappa_0}{\kappa}\right)^2 = \left(\frac{H_l}{H}\right)^2  = 1+ c\,\tilde{\eta}^{\frac{6(1+w)}{1+3w}}.
\end{equation}
It describes a smooth transition from exponential expansion at limiting curvature with $H \sim H_l$ to the late time stage dominated by the matter component with equation of state $w$ where $H \propto t^{-1}$.
The end of the inflationary stage, i.e. the end of accelerated expansion, happens at 
\begin{equation}
    \left(\frac{H_f}{H_l}\right)^2 = \frac{1+3w}{3(1+w)},
\end{equation}
or, in modified conformal time at
\begin{equation}
   \tilde{\eta}_f = \left(\frac{2}{(1+3w) c} \right)^{\frac{1+3w}{6(1+w)}}.
\end{equation}

%%%%%%%%%%%%%%%%%%%%%%%%%%%%%%%%%%%%%%%%%%%%%%%%%%%%%%%%%%%%%%%%%%%%%%
\section{Metric perturbations in comoving gauge}
\label{5:sec:perturbations}
%%%%%%%%%%%%%%%%%%%%%%%%%%%%%%%%%%%%%%%%%%%%%%%%%%%%%%%%%%%%%%%%%%%%%%

 In the same way as the modified Einstein equation of mimetic gravity takes its simplest form with the choice of time coordinate $t=\phi$,  the analysis of metric perturbations is most readily performed in comoving gauge $\delta \phi =0$.
Metric perturbations in different versions of mimetic gravity and also in different gauges have been analysed already a number of times, e.g. in 
\cite{Zheng:2017qfs,Ganz:2019vre,Takahashi:2017pje, Ijjas:2016pad, Firouzjahi:2017txv, HosseiniMansoori:2020mxj,Arroja:2015yvd, Gorji:2017cai,Cognola:2016gjy}. 
% \cite{Zheng:2017qfs}, \cite{Ganz:2019vre}, \cite{Takahashi:2017pje}, \cite{Ijjas:2016pad}, \cite{Firouzjahi:2017txv}, \cite{HosseiniMansoori:2020mxj}, \cite{Arroja:2015yvd}, \cite{Gorji:2017cai}, \cite{Cognola:2016gjy}
See appendix \ref{5:sec:appA} for the calculation of second order actions around a non-flat Friedmann background in a more general theory.

Starting with a general ADM metric,
\begin{equation}
    \textup{d}s^2 = N^2 \mathrm{d}t^2 - \gamma_{ij}\left (\mathrm{d}x^{i} +N^{i}\mathrm{d}t\right)\left (\mathrm{d}x^{j} +N^{j}\mathrm{d}t\right),
    \label{5:eq:ADM}
\end{equation}
the mimetic constraint fixes the lapse function as
\begin{equation}
N^{2}=\frac{\left(  \partial_{0}\phi-N^{i}\partial_{i}\phi\right)  ^{2}%
}{\left(  1+\gamma^{ij}\partial_{i}\phi\partial_{j}\phi\right)  }.\label{5:eq:lapse}%
\end{equation}
In comoving gauge where $\partial_{i}\phi=0$ this implies $N=\dot{\phi}$. Thus, the spatial slices coincide with slices of constant $\phi$ and the homogeneous lapse $N$ is determined only by the background. Using the background solution (\ref{5:eq:backgroundphi}) amounts to setting $N=1$. In this gauge it holds that the quantities
\begin{equation}
  \Box \phi = \kappa, \qquad \nabla^{\mu}\nabla^{\nu}\phi\nabla_{\mu}\nabla_{\nu}\phi = \kappa^{ij}\kappa_{ij}, \qquad \widetilde{R} ={^{3}\!R} 
\end{equation}
 are given in terms of the extrinsic curvature $\kappa_{ij} =\frac{1}{2}\left(\dot{\gamma}_{ij}-N_{i,j}-N_{j,i}\right)$ and intrinsic Ricci curvature $^3\!R_{ij}$ of spatial slices and their traces. 
 Similar, straightforward calculations show that the higher order terms from (\ref{5:eq:Lhd}) take the form
\begin{align}
     P^{\mu}_{\nu}\nabla_\mu\widetilde{R}\nabla^{\nu}\widetilde{R} &= -\gamma^{kl} \partial_{k}({^3\!R})\,\partial_{l}({^3\!R}),\\
    %  P^{\mu}_{\nu}P^{\alpha}_{\gamma}P^{\beta}_{\delta}\,\nabla_{\mu}\widetilde{R}^{\gamma\delta}\nabla^{\nu}\widetilde{R}_{\alpha\beta} &=  -\gamma^{kl}\bar{\nabla}_{k}({^3\!R^{ij}})\,\bar{\nabla}_{l}({^3\!R_{ij}})\\
\widetilde{C}^{\mu}_{\nu}\widetilde{C}^{\nu}_{\mu} &=  {^{3}C}^{i}_{j}\,{^{3}C}^{j}_{i}.
\end{align}
The spatial Cotton tensor is defined as in \cite{Chamseddine:2019gjh} by 
\begin{equation}
{^{3}C}_{j}^{i}=\frac{1}{\sqrt{\gamma}}\epsilon^{ikl}\bar{\nabla}_{k}\left(  {^{3}\!R}_{jl
}-\frac{1}{4}\gamma_{jl}\,{^{3}\!R}\right),
\end{equation}
where $\bar{\nabla}$ denotes the covariant derivative associated to $\gamma_{ij}$ and indices on spatial tensors are raised with $\gamma^{ij}$. 

Perturbing around the flat Friedmann background (\ref{5:eq:flatFriedmann}) in comoving gauge $\phi =t$, the metric perturbations can be further decomposed as
\begin{equation}
    \gamma_{ij}=a^2(t)\left(e^{-2\psi}{\delta_{ij}} - 2E_{,ij} -2F_{(i,j)} + (e^{h})_{ij}\right),\qquad \qquad N_{i} = \chi_{,i}-a S_{i}
\end{equation}
where
\begin{equation}
  F^{i}_{,i} = 0 , \qquad S^{i}_{,i}=0, \qquad h^{i}_{i} = 0, \qquad h^{i}_{j,i} = 0.
\end{equation}
Indices are raised with $\delta^{ij}$ and comma denotes partial derivative.
While the temporal coordinate is completely fixed, there is still a remaining freedom of choice in the spatial coordinates. In contrast to the synchronous gauge condition $N_{i} =0$, the choice $E=0$, $F_{i}=0$ fixes the coordinates uniquely, cf. \cite{Mukhanov}.

%%%%%%%%%%%%%%%%%%%%%%%%%%%%%%%%%%%%%%%%%%%%%%%%%%%%%%%%%%%%%%%%%%%%%%
\subsection{Vector perturbations}
%%%%%%%%%%%%%%%%%%%%%%%%%%%%%%%%%%%%%%%%%%%%%%%%%%%%%%%%%%%%%%%%%%%%%%
In mimetic gravity, like in GR, the vector perturbations parametrized by $F_i$ and $S_i$ as
\begin{equation}
    \gamma_{ij}=a^2(t)\left({\delta_{ij}} -F_{i,j}-F_{j,i}\right),\qquad \qquad N_{i} = -a S_{i},
\end{equation}
are non-dynamical in the absence of sources. 
Note that the spatial Ricci tensor $^{3}\!R_{ij}= \mathcal{O}(F^2)$ is second order in vector perturbations when perturbing around a spatially flat Friedmann universe.
Hence, higher order spatial curvature terms like in (\ref{5:eq:Lhd}) do not contribute to the linearized equation of motion for vector perturbations.
Choosing the gauge $S_{i}=0$, we can use the equations of motion derived in synchronous coordinates in \cite{Chamseddine:2019fog} and express them in terms of the gauge invariant variable $V_{i} = S_{i}-a\dot{F}_{i}$.
Assuming hydrodynamical matter of the perfect fluid type, the only non-vanishing vector components of the perturbed energy momentum tensor $\delta T_{\mu\nu}$ are of the form $\delta T_{0i} =\left(\varepsilon^{m}+p\right) \delta u_{\perp i} $, \cite{Mukhanov}.

The spatial modified Einstein equations become
\begin{equation}
   \partial_t \left(a^2 f \left(  V_{i,j}+V_{j,i}\right) \right)=0,
\end{equation}
and have the solution
\begin{equation}
    V_i = \frac{C_{\perp i} }{f a^2}.
\end{equation}
Note that the gauge invariant vector perturbation $V_i$ decays both in the late time limit, like in GR, as well as in the early time limit, as $f\to \infty$. This is completely in line with the previous results from \cite{Chamseddine:2019fog} that due to asymptotic freedom anisotropies decay during contraction at limiting curvature. Provided that the constant of integration $C_{\perp i}$ and thus the maximum value of $V_i$ is bounded, we see that in this model vector perturbations never become important.
Reinserting this solution into the $0-i$ modified Einstein equation
\begin{equation}
    f \Delta V_i = 16\pi \left(\varepsilon^{m}+p\right)a\, \delta u_{\perp i}, 
\end{equation}
with $\Delta = \delta^{kl}\partial_k\partial_l$,
shows that for the physical velocity $\delta v^{i}= -a^{-1}  \delta u_{\perp i} $ it holds that
\begin{equation}
   \delta v^{i} = -\frac{\Delta C_{\perp i}}{16\pi a^4 \left(\varepsilon^{m}+p\right) },
   \label{5:eq:deltavi}
\end{equation}
where $f$ cancels and which is thus the same equation that one finds in standard GR. Note that only the matter energy density and pressure are appearing since mimetic matter does not source vector perturbations.

\clearpage
%%%%%%%%%%%%%%%%%%%%%%%%%%%%%%%%%%%%%%%%%%%%%%%%%%%%%%%%%%%%%%%%%%%%%%
\subsection{Tensor perturbations}
%%%%%%%%%%%%%%%%%%%%%%%%%%%%%%%%%%%%%%%%%%%%%%%%%%%%%%%%%%%%%%%%%%%%%%
Tensor perturbations are parametrized by
\begin{equation}
    \gamma_{ij} = a^2\left(\eta_{ij}+h_{ij}+\frac{1}{2}h_{i}^{k}h_{kj}\right), \qquad N_{i} = 0,
\end{equation}
where indices on $h_{ij}$ are raised with $\delta^{ij}$ and 
$ h^{i}_{i}=0$, $h^{i}_{j,i} =0$.
Expanding (\ref{5:eq:2Szeta}) to second order in $h_{ij}$ yields the second order action (see appendix \ref{5:sec:appA})
\begin{equation}
{^{(2)}\mathcal{S}}_{h} = \frac{1}{64\pi}\int\!\mathrm{d}t\,\mathrm{d}^3x\, a^3  \bigg \{  f\,(\dot{h}_{ij})^2- \frac{1}{a^2}(\partial h_{ij})^2 -\frac{\sigma_T^4}{a^6}(\partial^3 h_{ij})^2\bigg \},
\end{equation}
where $\dot{}$ denotes $t$ derivatives, $(\partial h_{ij})^2 \equiv \delta^{kl}\partial_k h^{ij}\partial_l h_{ij}$ and  $(\partial^3 h_{ij})^2 \equiv (\partial \Delta h_{ij})^2$.
Since $f>0$, tensor perturbations do not exhibit any instabilities provided that $\sigma_T^4\geq0$. 

We can read off the propagation speed of gravitational waves, 
\begin{equation}
    c_{T}^2 = f^{-1}(\kappa) = G(\kappa),
\end{equation}
and find that it deviates from unity in the early time/high curvature regime. In particular, note that $c_T$ is vanishing at limiting curvature $\kappa = \kappa_0$ in the case of asymptotic freedom. Using (\ref{5:eq:f}), right at the end of inflation $c_{T}^2 =\tfrac{2}{3(1+w)}$. Already one second later, in the late time regime where $\kappa \propto t^{-1}$ and $t$ roughly corresponds to the time since the end of inflation, the propagation speed of gravitational waves is approximately
\begin{equation}
    1-c_{T} \approx1- \sqrt{1-\left(\frac{1}{ \kappa_{0}\,t }\right)^2} \lesssim \left(\frac{\kappa_{pl}}{\kappa_{0}}\right)^2\left(\frac{1\, \textup{sec}}{t}\right)^2  \times 10^{-86}.
\end{equation}
The limiting curvature $\kappa_0$ could naturally be taken to lie a few orders of magnitude below the Planck curvature $\kappa_{pl} = 1/t_{pl}$ for the sake of singularity resolution.
Thus, late time experimental constraints like $1-c_T < \mathcal{O}(10^{-15})$, which stems from the multi messenger event GW170817 (\cite{TheLIGOScientific:2017qsa}, \cite{Monitor:2017mdv})  that happened around $t \sim 10^{9}$ years after the conjectured inflationary period, are not touched in the slightest by this deviation of $c_T$ from the speed of light in the limiting curvature regime. 
They do, however, tightly constrain low curvature / IR modifications due to additional terms in the Lagrangian like $(\Box\phi)^2$ or in particular $\nabla^{\mu}\nabla^{\nu}\phi\nabla_{\mu}\nabla_{\nu}\phi$, see \cite{Casalino:2018wnc}, cf. \cite{Gumrukcuoglu:2017ijh}.

\paragraph{Mode expansion.}
Introducing $y = a f^{1/4}$ and substituting the expansion
\begin{equation}
    h^i_{j}(\tilde{\eta},\textbf{x}) = \sqrt{32\pi}\int\! \frac{\textup{d}^3 k}{(2\pi)^{3/2}} \frac{u_{k}(\tilde{\eta})\,  e^{i \textbf{k} \textbf{x}}}{y\sqrt{\mathrm{e}^m_{n}\mathrm{e}^n_{m}}} \,  \mathrm{e}^i_{j}(\textbf{k}) , 
\end{equation}
where the mode function $u_k$ and the contraction $\mathrm{e}^m_{n}\mathrm{e}^n_{m}$ of the polarization tensor only depend on the magnitude of $\textbf{k}$ by isotropy,
the second order action in modified conformal time $\mathrm{d} \tilde{\eta} = \frac{\mathrm{d}t}{a\sqrt{f}}$ becomes
\begin{equation}
     {^{(2)}\mathcal{S}}_{h}  = \frac{1}{2} \int \!\textup{d}\tilde{\eta}\,\textup{d}^3 k \left\{u_{k,\tilde{\eta}}^2-\left(k^2+\frac{\sigma_T^4}{a^4}k^6-\frac{y_{\tilde{\eta}\tilde{\eta}}}{y}\right)u_{k}^2\right\}.
\end{equation}
A subscript $u_{k,\tilde{\eta}} \equiv  \partial u_{k} /\partial \tilde\eta $ denotes derivatives. The modified Mukhanov-Sasaki equation for $u_k$ reads
\begin{equation}
    u_{k,\tilde{\eta}\tilde{\eta}} + \left(\frac{\sigma_T^4}{a^4}k^6+k^2-\frac{y_{\tilde{\eta}\tilde{\eta}}}{y}\right)u_{k} = 0,
    \label{5:eq:u-modeeq}
\end{equation}
and the power spectrum of gravitational waves is given by \cite{Mukhanov}
\begin{equation}
    \delta^2_{h}(k,\tilde{\eta}) = \frac{8}{\pi} \frac{k^3 \left|u_{k}\right|^2 }{y^2}.
\end{equation}
In the late time limit $f\to1$, $\tilde{\eta} \to \eta$ is identical to regular conformal time and (\ref{5:eq:u-modeeq}) becomes just like in GR,
\begin{equation}
    u_{k,\eta\eta} + \left(k^2-\frac{a_{\eta\eta}}{a}\right)u_{k} = 0.
\end{equation}
Initially, at $\tilde{\eta}\to 0$ the $k^6$ term is dominating.
The requirement of an initial minimal level of quantum fluctuations determines the initial conditions at $\tilde{\eta} \to 0$ (up to an unimportant phase factor, see \cite{Mukhanov}) as 
\begin{equation}
    u_{k} = \frac{a}{\sigma_T k^{3/2} }, \qquad u_{k,\tilde{\eta}} = i\, \frac{\sigma_T k^{3/2}}{a}.
    \label{5:eq:u-IC}
\end{equation}
In the initial region where the $k^6$ term is dominating and $a_{\tilde{\eta}}\, a \ll k^3 \sigma_T^2$, the solution of (\ref{5:eq:u-modeeq}) with initial conditions (\ref{5:eq:u-IC}) is well described by the WKB approximation
\begin{equation}
    u_k = \frac{a(\tilde{\eta})}{\sigma_T k^{3/2} } \exp\left(-i \int\!\mathrm{d} \tilde{\eta} \frac{\sigma_T^2 k^3}{a(\tilde{\eta})^2}\right).
\end{equation}
Note that the initial spectrum is scale invariant, by virtue of the same higher order $k^6$ term that is needed for a power counting renormalizable theory.
After ``horizon'' exit, i.e. when $y_{\tilde{\eta}\tilde{\eta}}/y$ becomes dominating, the solution of (\ref{5:eq:u-modeeq}) is
\begin{equation}
    u_{k}(\tilde{\eta}) = y(\tilde{\eta})  \left(A^{T}_{k}+ B^{T}_{k}  \int^{\infty}_{\tilde{\eta}}\!\frac{\mathrm{d}\tilde{\eta}'}{y^2(\tilde{\eta}')} \right),
    \label{5:eq:u-superh}
\end{equation}
where the second term is decaying compared to the first term and will be ignored for the following estimates. The primordial spectrum of tensor perturbations is hence given by
\begin{equation}
        \delta^2_{h}(k,\tilde{\eta}) \approx \frac{8 }{\pi}k^3 \left|A^{T}_{k}\right|^2.
\end{equation}
The initially scale invariant spectrum would only be preserved for modes which exit the ``horizon'' before the $k^2$ term starts do dominate and in a region where $y \propto a$, i.e. where the propagation speed $c_T$ is almost constant. This is certainly not satisfied for modes exiting the horizon during the inflationary stage of an asymptotically free model where $f$ is rapidly changing. Hence we can expect a primordial spectrum of tensor perturbations which is far from being scale invariant on large scales.

\paragraph{Radiation dominated background.}
For a concrete example with the simplest possible background evolution, consider the case where the total energy density is dominated by a component with equation of state $w=1/3$. In this case, fixing $c$ by setting $a(\tilde{\eta}_f)=1$ at the end of inflation, the solution (\ref{5:eq:mfsolution}) is $a(\tilde{\eta}) = H_l \tilde{\eta}$ and we find
\begin{equation}
    y = \sqrt[4]{1+(H_l\tilde{\eta})^4}, \qquad \frac{y_{\tilde{\eta}\tilde{\eta}}}{y} = \frac{3H_l^4 \tilde{\eta}^2}{(1+(H_l\tilde{\eta})^4)^2}.
\end{equation}
Before the end of inflation at $\tilde{\eta}_f = 1/H_l$, the modified Mukhanov-Sasaki equation (\ref{5:eq:u-modeeq}) is well approximated by
\begin{equation}
    u_{k,\tilde{\eta}\tilde{\eta}} + \left(\frac{\tilde{\sigma}_T^4 }{H_l^8 \tilde{\eta}^4}k^6+k^2-3H_l^4 \tilde{\eta}^2\right)u_{k} = 0,
    \label{5:eq:u-modeeq2}
\end{equation}
with the dimensionless $\tilde{\sigma}_T := \sigma_T H_l$. The ``horizon'' exit of the mode $k$ happens at $\tilde{\eta}_{k} =  k/H_l^2\sqrt{ D}$ where $D$ solves
\begin{equation}
    \tilde{\sigma}_T^4 D^3+ D =3.
\end{equation}
Note that by the assumption made before, this is only valid for modes $k^2\ll D H_l^2$ which exit the ``horizon'' before the end of inflation, i.e. $\tilde{\eta}_k \ll \tilde{\eta}_f$.

In the initial region where the $k^6$ term is dominating, the solution of (\ref{5:eq:u-modeeq2}) is given by 
\begin{equation}
    u_k(\tilde{\eta})  = \frac{H_l^2\,\tilde{\eta} }{ k^{3/2}\tilde{\sigma}_T} \exp  \left(-i\frac{\tilde{\sigma}_{T}^2 k^3}{H_l^4 \tilde{\eta}} \right),
    \label{5:eq:u-initialsol2}
\end{equation}
where the initial conditions (\ref{5:eq:u-IC}) have been taken into account. 

It depends only on $\tilde{\sigma}_T$ if modes will exit the ``horizon'' before or after the gradient term is dominating over the $k^6$ term.
For $\tilde{\sigma}_T \gg 1$ the gradient term will never become important and $D=(3/\tilde{\sigma}_T^4)^{1/3}$. By matching the absolute value of $\left|{u_k}\right|$ from (\ref{5:eq:u-initialsol2}) and (\ref{5:eq:u-superh}) we can estimate
\begin{equation}
    \left|A^{T}_{k}\right| \approx  \frac{H_l}{ k^{3/2}\tilde{\sigma}_T} \frac{a(\tilde{\eta}_k)}{y(\tilde{\eta}_k)}  \approx \frac{1}{ \sqrt{D}\, \tilde{\sigma}_T\,k^{1/2}}.
\end{equation}

On the other hand, if $\tilde{\sigma}_T \ll1$ then there is an intermediate region where the $k^2$ term is dominating and $D=3$. In this region, starting at
\begin{equation}
    \tilde{\eta}_\ast = \frac{\tilde{\sigma}_{T}}{H_l^2} \,k < \tilde{\eta}_k = \frac{k}{\sqrt{3}H_l^2},
\end{equation}
the absolute value $\left|u_k\right|$ is approximately constant and we can estimate
\begin{equation}
    \left|A^{T}_{k}\right| \approx  \frac{H_l}{ k^{3/2}\tilde{\sigma}_T} \frac{a(\tilde{\eta}_\ast)}{y(\tilde{\eta}_k)}  \approx \frac{1}{k^{1/2}}.
\end{equation}
Note that this is identical to the result that one would get if $\tilde{\sigma}_T =0$ and the initial conditions were determined from the $k^2$ term.

In summary, the primordial spectrum of large wavelength modes $k^2\ll D H_l^2 $ becomes
\begin{equation}
    \delta^2_{h}(k,\tilde{\eta}) \approx 
    \begin{cases}  
\frac{8 }{\pi}\, k^2  & \text{ if } \tilde{\sigma}_T \ll 1\\
\frac{8}{\pi}\, \frac{k^2}{(\sqrt{3}\,\tilde{\sigma}_T)^{2/3}} 
 & \text{ if } \tilde{\sigma}_T \gg 1 
\end{cases}
\label{5:eq:h-spectrum}
\end{equation}
Note that here $1/k$ is the physical wavelength at the end of inflation.
In both cases the spectrum is indeed far from being scale invariant, with a large blue tilt and a spectral index of
\begin{equation}
    n_T = 2.
\end{equation}

For a different background, e.g. $w=0$ the situation becomes more complicated because $y_{\tilde{\eta}\tilde{\eta}}/y$ initially diverges as $\sim-1/\tilde{\eta}^2$. Hence, after an initial sub-``horizon'' $k^6$ domination region, there can be an earlier intermediate super-``horizon'' region, followed again by a sub-``horizon'' region where $k^2$ dominates before finally exiting the horizon. A similar situation occurs for scalar perturbations in a dust dominated background, see below.

%%%%%%%%%%%%%%%%%%%%%%%%%%%%%%%%%%%%%%%%%%%%%%%%%%%%%%%%%%%%%%%%%%%%%%
\subsection{Scalar perturbations}
%%%%%%%%%%%%%%%%%%%%%%%%%%%%%%%%%%%%%%%%%%%%%%%%%%%%%%%%%%%%%%%%%%%%%%
The scalar metric perturbations in comoving gauge are characterized through $\zeta$ and $\chi$ by
\begin{equation}
    \gamma_{ij}=a^2(t)e^{2\zeta}\delta_{ij},\qquad\qquad  N_{i} = \chi_{,i},
\end{equation}
where 
\begin{equation}
    \zeta = -\psi - \frac{H}{\dot{\phi}} \delta \phi .
\end{equation}
is the gauge invariant curvature perturbation in comoving gauge $\delta \phi =0$.

In contrast to standard GR, even in the absence of any matter fluctuations the conformal degree of freedom of mimetic gravity can become dynamical. In this case the action (\ref{5:eq:2Szeta}) expanded to second order in scalar perturbations (see appendix \ref{5:sec:appA}) becomes
\begin{equation}
    \begin{aligned}
{^{(2)}\mathcal{S}} = \frac{1}{8 \pi}
\int\! \mathrm{d}^4x\sqrt{\eta} \, a^{3}\bigg{\{}&
     -3\ell''\dot{\zeta}^2+ \tfrac{1}{a^2}(\partial \zeta)^2 -\tfrac{\sigma_S^4}{a^6}\,(\partial^3 \zeta)^2 
      \phantom{\bigg{\}}}\\&
   - \left[ \tfrac{1}{3}(\ell''-f)\Delta \chi -2\ell''a^2 \dot{\zeta}\right]\tfrac{1}{a^4}\Delta\chi
     \bigg{\}},
\end{aligned}
\end{equation}
where $\dot{}$ denotes $t$ derivatives, $(\partial \zeta)^2 \equiv \delta^{kl}\partial_k \zeta \partial_l \zeta$ and  $(\partial^3 \zeta)^2 \equiv (\partial \Delta \zeta)^2$.
In the case $\ell''-f = 0$, variation with respect to $\chi$ shows that $\dot{\zeta}=0$ and thus the conformal degree of freedom is frozen.
In the case $\ell''-f \neq 0$, the second order action after integrating out $\chi$ becomes
\begin{equation}
{^{(2)}\mathcal{S}_{\zeta}}= \frac{1}{8\pi}
\int\!\mathrm{d}t\,\mathrm{d}^3x\, a^{3}\bigg{\{}
    \frac{3 \ell'' f }{\ell''-f}\,\dot{\zeta}^2+ \frac{1}{a^2}\left( \partial\zeta\right)^2-\frac{\sigma_{S}^4}{a^6}\left( \partial^3\zeta\right)^2  \bigg{\}}.
   \label{5:eq:2Szeta}
\end{equation}

\paragraph{No ghost instability.}
Since $f>0$, the condition to have no ghost instability reads
\begin{equation}
    \frac{\ell ''}{\ell''-f}>0\qquad \Rightarrow  \quad \ell'' > f.
    \label{5:eq:ghostfree1}
\end{equation}
The other possible case $\ell'' <0$ was excluded because for a smooth low curvature GR limit it must hold that $\ell''\to 1$, $f\to1$ as $\kappa \to 0$.

The condition (\ref{5:eq:ghostfree1}) constrains the slope of the modified Friedmann equation (\ref{5:eq:mFriedmann}) by
\begin{equation}
 \frac{8 \pi}{3} \frac{\partial \varepsilon }{\partial {H^2}}  = \frac{1}{\kappa} \frac{\partial}{\partial {\kappa}}  \left(\kappa\ell'-\ell\right) = \ell'' \overset{!}{>} f > 0,
 \label{5:eq:ghostfree2}
\end{equation}
which can be rewritten as
\begin{equation}
    0< \frac{\partial H^2}{\partial \varepsilon}< \frac{8 \pi}{3} G(\kappa).
    \label{5:eq:ghostfree3}
\end{equation}
Note that this condition is fully general and does not make use of the simplifying choice of $\Lambda$ given by (\ref{5:eq:simplifying}).
Using (\ref{5:eq:simplifying}), the condition becomes $\kappa G' <0$ and we see that the running gravitational constant can only decrease when going to higher curvatures. This condition is always satisfied for the asymptotically free background solutions from section \ref{5:sec:background}.

\paragraph{Instability of spatially flat bouncing solutions.}
In a previous mimetic model \cite{Chamseddine:2016uef}, limiting curvature was realized by a bounce. In order to achieve a bouncing background solution in a spatially flat universe, the modified relation $a(H^2)$ cannot be one-to-one. 
In fact, the generic background evolution described in section \ref{5:sec:background} can only be circumvented by including multi-valued functions with intricate branch changes in the Lagrangian, cf. \cite{Brahma:2018dwx}.
Moreover, it was discussed in \cite{deCesare:2019pqj} that the bouncing solution from \cite{Chamseddine:2016uef} is unstable under perturbations.
From the above it is easy to see that this is an unavoidable feature of any bouncing solution of (\ref{5:eq:mFriedmann}): In order to obtain a bounce there must be a region where $H^2(\varepsilon)$ decreases until it eventually reaches a zero at some finite $a_{\min}> 0$. In this region the condition (\ref{5:eq:ghostfree3}) is violated and hence there is a ghost instability.

The stability analysis for bounces in non-flat universes found in \cite{Chamseddine:2019fog}, driven by higher order spatial curvature terms, is performed in appendix \ref{5:sec:appB}.

\paragraph{Gradient instability?}
As has been noticed in \cite{Ijjas:2016pad}, mimetic models without any higher derivatives typically exhibit a gradient instability. In fact, also for models which include higher derivatives, a negative square of the speed of sound $c_S^2 <0$ is something we have to deal with, at least in the low curvature regime, in any mimetic model with a well behaved GR limit. However, in this same limit also $\ell'' \to f$ and thus the scalar degree of freedom of mimetic gravity stops to propagate. 

In a theory with higher mixed derivatives, the sign of the gradient term can be made negative in the high curvature regime. In fact, for the Lagrangian (\ref{5:eq:Lhmd}) the second order action is
\begin{align}
{^{(2)}\widetilde{\mathcal{S}}_{\zeta}} = \frac{1}{8\pi}
\int\!\mathrm{d}t\,\mathrm{d}^3x\, a^{3}\bigg{\{}
   & \frac{3 \ell''f}{\ell''-f}\,\dot{\zeta}^2+\left[f +  \tfrac{1}{a} \partial_{0}\left(a\frac{6f}{\kappa}\right)\right]  \tfrac{1}{a^2}(\partial \zeta)^2+ \frac{6f'}{\kappa}\tfrac{1}{a^4}\left(\partial^2\zeta\right)^2\bigg{\}}. \label{5:eq:2Shmd}
\end{align}
A similar second order action has been used in \cite{Zheng:2017qfs} and in \cite{HosseiniMansoori:2020mxj}. However, the models considered in these works  were not symmetric under shifts of $\phi$. Using a $\phi$ dependent potential essentially amounts to the introduction of a time dependent background. In this way it is  easy to produce any background evolution one could wish for, including an inflationary stage, cf. \cite{Chamseddine:2014vna}. In such a highly flexible model, indeed the sign of the gradient term in (\ref{5:eq:2Shmd}) can be adjusted more or less independent of the background. However, note that such an adjustment must restrict to the high curvature regime. Otherwise the GR limit is violated, as it happens in \cite{HosseiniMansoori:2020mxj}.

Sticking to shift-symmetric mimetic models, the background evolution and the evolution of perturbations are no longer decoupled, but must be driven by the same dynamics.
Trying to use asymptotic freedom and the background solution from section \ref{5:sec:background} in the model (\ref{5:eq:2Shmd}), we find that the gradient term would actually become negative during the inflationary stage, however the sign of the $(\partial^2\zeta)^2$ term is strictly positive and diverges as $f'\to \infty$. This shows that even though higher mixed derivatives can make $c_S^2$ positive at high curvatures, in shift symmetric models they actually lead to a higher order instability that could be removed only by a high amount of tuning.

Instead, let us return to our original theory (\ref{5:eq:S}) without higher mixed derivatives but with higher spatial derivatives coming from (\ref{5:eq:Lhd}). In the corresponding second order action (\ref{5:eq:2Szeta}) the gradient term comes with the constant prefactor $+1$ and in the case $\sigma_S=0$ we would read off the speed of sound
\begin{equation}
    c_S^2 = - \frac{\ell''-f}{3 \ell'' f} <0.
\end{equation}
We will continue to use the name $c_S^2$ to refer to this quantity also in the case $\sigma_S \neq 0$, but it is important to note that in this case due to the modified dispersion relation the true speed of sound is in general different from $c_S^2$.
Note that $c_S^2$ has to be negative throughout by (\ref{5:eq:ghostfree1}). Using (\ref{5:eq:f}), it can be rewritten as
\begin{equation}
    c_S^2 = -\frac{G(1-G)}{3}.
\end{equation}
Note that $c_S^2$ goes to zero in the late time limit as required by the GR limit $G\to 1$, but it also goes to zero in the early time limit as $G\to 0$. The minimal value $c_S^2 = -1/12$ is reached when $G=1/2$ at $\kappa = \kappa_0/\sqrt{2}$. At the earliest times, the gradient term in (\ref{5:eq:2Szeta}) will be dominated by the higher order spatial derivative term $\tfrac{1}{a^6} (\partial^3\zeta)^2$. In this region there is no instability, provided that $\sigma_S^4>0$.
Thus, the potential gradient instability region is sandwiched between the region of domination of the higher order term and the late time region where the scalar degree of freedom is ``frozen''. These two other regions are without instabilities and hence the gradient instability gets to act, if at all, only during a limited time. As I will show below in two concrete examples, provided that $\sigma_S \gg 1/H_l$, the wrong sign gradient term cannot lead to any dangerous instability.

\paragraph{Mode expansion.}
Introducing the time coordinate $\tau$ and the expression $z$,
\begin{equation}
\mathrm{d}t= a \sqrt{f}\, \mathrm{d}\tilde{\eta} = a\,\sqrt{\frac{3f\ell''}{\ell''-f}} \, \mathrm{d}\tau, \qquad 
z = \frac{a}{\sqrt{4\pi}} \left(\frac{3f\ell''}{\ell''-f}\right)^{1/4},
\label{5:eq:tauz}
\end{equation}
the second order action (\ref{5:eq:2Szeta}) written in terms of the canonically normalized variable $v = z \zeta$ becomes
\begin{equation}
 {^{(2)}\mathcal{S}_{\zeta}}=
 \frac{1}{2} \int\! \mathrm{d}\tau\,\mathrm{d}^3x
  \left\{v_\tau^2+(\partial v)^2- \frac{\sigma_S^4}{a^4} (\partial^3 v)^2+\frac{z_{\tau\tau}}{z} v^2\right\}.
\end{equation}
A subscript $v_{\tau} \equiv  \partial v /\partial \tau$ denotes derivatives. Performing the mode expansion of $v$ into Fourier modes $v_k$, the modified Mukhanov-Sasaki equation becomes
\begin{equation}
    v_{k,\tau\tau}+\left(\frac{\sigma_S^4}{a^4} k^6- k^2-\frac{z_{\tau\tau}}{z}\right)v_{k}=0.
    \label{5:eq:v-modeeq}
\end{equation}
The power spectrum of the curvature perturbation $\zeta$ is given by \cite{Mukhanov}
\begin{equation}
    \delta^2_{\zeta}(k) = \frac{k^3 \left|v_{k}\right|^2}{2 \pi^2 z^2} .
\end{equation}
In the initial region where the $k^6$ term is dominating and $a_{\tilde{\eta}}\, a \ll k^3 \sigma_S^2$, the solution of (\ref{5:eq:v-modeeq}) with quantum vacuum initial conditions is well described by the WKB approximation
\begin{equation}
    v_k = \frac{a(\tilde{\tau})}{\sigma_S k^{3/2} } \exp\left(-i \int\!\mathrm{d} \tau \frac{\sigma_S^2 k^3}{a(\tilde{\eta})^2}\right),
\end{equation}
and it has a scale invariant spectrum.
After ``horizon'' exit, when $z_{\tau\tau}/z$ is dominating, the solution of (\ref{5:eq:v-modeeq}) is
\begin{equation}
    v_{k}(\tau) = z(\tau)  \left(A^{S}_{k}+ B^{S}_{k}  \int^{\infty}_{\tau}\!\frac{\mathrm{d}\tau'}{z^2(\tau')} \right),
    \label{5:eq:v-sol-superhorizon}
\end{equation}
where the second term is decaying compared to the first term and will be ignored in the following estimates. At late times the primordial spectrum after horizon exit is
\begin{equation}
    \delta_\zeta^2 \approx \frac{k^3 }{2 \pi^2} \left|A^S_{k}\right|^2.
\end{equation}
The wrong sign gradient term $-k^2$ will never get to dominate and cannot cause instability, provided that at ``horizon'' exit the condition
\begin{equation}
    \frac{\sigma_S^4}{a^4}k^6 \sim \frac{z_{\tau\tau}}{z}\gg k^2
    \label{5:eq:noinstability-condition}
\end{equation}
holds. In other words, the physical wavelength $\lambda_{\textup{phys}} = a/k$ at horizon exit should satisfy
\begin{equation}
    \lambda_{\textup{phys}}\bigg | _{\frac{\sigma_S^4}{a^4}k^6 \sim \frac{z_{\tau\tau}}{z}} \ll \sigma_S.
\end{equation}
The initial scale invariant spectrum would be preserved only if at horizon exit $z\propto a$, i.e.
\begin{equation}
  -\frac{1}{c_S^2}=  \frac{3f\ell''}{\ell''-f} \approx const. 
    \label{5:eq:csconst}
\end{equation}
This condition will in general not be satisfied both at late times where $\ell''-f \to 0$ and at early times where $f\to \infty$, $\ell''\to \infty$.
Thus, without a substantial amount of tuning of the functions $f$ and $\Lambda$ in order for (\ref{5:eq:csconst}) to be satisfied at least in some intermediate region where the relevant modes exit the ``horizon'', one can already expect a primordial spectrum that will be far from scale invariant.

Note that all modes $\zeta_k$ will at some point exit the ``horizon'' and never again re-enter, also after transitioning to the post-inflationary phase. This is a manifestation of the fact that in the GR limit the scalar degree of freedom of pure mimetic gravity is dust-like and non-propagating. 
It is clear that aiming for a more realistic model, matter perturbations would have to be taken into account.

\paragraph{Radiation dominated background.}
In modified conformal time $\tilde{\eta}$ the radiation dominated background solution from section \ref{5:sec:background} is given by $a(\tilde{\eta}) = H_l \tilde{\eta}$, where $c$ was fixed by setting $a(\tilde{\eta}_f) = 1$, and we can find the explicit expressions
\begin{equation}
z= \sqrt{\tfrac{\sqrt{3}}{4\pi}\left(1+(H_l\tilde{\eta})^4\right)}, \qquad 
    \frac{ \mathrm{d} \tilde{\eta}}{\mathrm{d} \tau} = \sqrt{3(1+(H_l\tilde{\eta})^4)},\qquad 
    \frac{z_{\tau\tau}}{z} = 18H_l^4 \tilde{\eta}^2.
\end{equation}
The modified Mukhanov-Sasaki equation (\ref{5:eq:v-modeeq}) becomes
\begin{equation}
    v_{k,\tau\tau}+\left(\frac{\tilde{\sigma}_S^4}{H_l^8 \tilde{\eta}^4} k^6 - k^2-18H_l^4 \tilde{\eta}^2\right)v_{k}=0,
    \label{5:eq:v-modeeq-radiation}
\end{equation}
where the dimensionless $\tilde{\sigma}_S = H_l \sigma_S$ was introduced. 
Assuming that $\tilde{\sigma}_S \gg \mathcal{O}(1)$, the ``horizon'' crossing happens at
\begin{equation}
    \tilde{\eta}_k = \left(\frac{\tilde{\sigma}_S^4}{18}\right)^{1/6} \frac{k}{H_l^2}
\end{equation}
and the condition (\ref{5:eq:noinstability-condition}) is satisfied for all modes.

Accelerated expansion ends at $\tilde{\eta}_f = 1/H_l$ and before that we can approximate $\tau = \tilde{\eta}/\sqrt{3}$ and write the $\tau$ derivatives in (\ref{5:eq:v-modeeq-radiation}) as $\tilde{\eta}$ derivatives. Matching the absolute value of the initial solution with quantum vacuum initial conditions
\begin{equation}
    v_k(\tilde{\eta})  = \frac{H_l^2\,\tilde{\eta} }{ 3^{1/4}\tilde{\sigma}_S k^{3/2}} \exp  \left(-i\frac{\sqrt{3}\tilde{\sigma}_{S}^2 k^3}{H_l^4 \tilde{\eta}} \right),
    \label{5:eq:u-initialsol}
\end{equation}
 to the solution (\ref{5:eq:v-sol-superhorizon}) after horizon exit, we can estimate
the late time spectrum of large wavelength modes $k^2\ll H_l^2 (18/\tilde{\sigma}_S^4)^{1/3}$ which exit the horizon before the end of inflation as
\begin{equation}
   \delta^2_{\zeta}(k) \approx  \frac{(2/3)^{2/3}}{3 \pi} \frac{k^2}{\tilde{\sigma}_S^{2/3}}.
   \label{5:eq:ScalarSpectrum}
\end{equation}
Note that here $1/k$ is the physical wavelength at the end of inflation.
It is far from being scale invariant with a large blue tilt and a spectral index of 
\begin{equation}
    n_S-1 =2.
\end{equation}
Combining with (\ref{5:eq:h-spectrum}), the tensor to scalar ratio is given by
\begin{equation}
    r = \frac{\delta^2_h}{\delta^2_\zeta} \approx
    \begin{cases}24 \left(\tfrac{3}{2}\tilde{\sigma}_S\right)^{2/3}
 & \text{ if } \tilde{\sigma}_T \ll 1 \\ 
24\left(\frac{\sqrt{3}}{2} \frac{\tilde{\sigma}_S}{\tilde{\sigma}_T}\right)^{2/3}& \text{ if } \tilde{\sigma}_T \gg 1 
\end{cases}
\end{equation}
It can be small only if $\sigma_T \gg \sigma_S \gg 1/H_l$.

\paragraph{Dust/MDM dominated background.}
If we consider now the background solution dominated by dust or mimetic dark matter with equation of state $w=0$, the situation gets complicated by an additional early intermediate super-``horizon'' region. Even though in this case long wavelength modes go through a short gradient instability phase, we will find that the growth of modes during this stage is completely negligible if $\sigma_S \gg 1/H_l$.

Using the background solution in modified conformal time $\tilde{\eta}$ from section \ref{5:sec:background} with $w=0$ and fixing $c$ by $a(\tilde{\eta}_f)=1$ at the end of inflation, the scale factor is given by $a(\tilde{\eta}) = H_l^2\tilde{\eta}^2 / 8$ and we calculate
% $a(\tilde{\eta}) = 2 \sqrt{c}\, \tilde{\eta}^2 / H_l$ and we calculate
% \begin{equation}
%         z= \frac{(3c)^{1/4}}{\sqrt{\pi}H_l} \sqrt{\tilde{\eta}\left(1 + c \tilde{\eta}^6\right)}, \qquad 
%     \frac{ \mathrm{d} \tilde{\eta}}{\mathrm{d} \tau} = \sqrt{3(1+c\tilde{\eta}^6)} ,\qquad 
%     \frac{z_{\tau\tau}}{z} =\frac{3 (-1 + 77 c \tilde{\eta}^6)}{4 \tilde{\eta}^2}.
% \end{equation}
% \begin{equation}
%         z= \frac{(3)^{1/4}}{\sqrt{16\pi}} \sqrt{H_l\tilde{\eta}\left(1 + 2^{-8}(H_l \tilde{\eta})^6\right)}, \qquad 
%     \frac{ \mathrm{d} \tilde{\eta}}{\mathrm{d} \tau} = \sqrt{3(1+2^{-8}(H_l\tilde{\eta})^6)} ,\qquad 
%     \frac{z_{\tau\tau}}{z} =\frac{3 (-1 + 77\times2^{-8} (H_l\tilde{\eta})^6)}{4 \tilde{\eta}^2}.
% \end{equation}
% \begin{equation}
%         z= \frac{3^{1/4}}{\sqrt{16\pi}} \sqrt{H_l\tilde{\eta}\left(1 + 2a^3\right)}, \qquad 
%     \frac{ \mathrm{d} \tilde{\eta}}{\mathrm{d} \tau} = \sqrt{3(1+2a^3)} ,\qquad 
%     \frac{z_{\tau\tau}}{z} =\frac{3 (-1 + 154 a^3)}{4 \tilde{\eta}^2}.
% \end{equation}
\begin{equation}
        z=  \sqrt{\tfrac{\sqrt{3}}{16\pi}H_l\tilde{\eta}\left(1 + 2a^3\right)}, \qquad 
    \frac{ \mathrm{d} \tilde{\eta}}{\mathrm{d} \tau} = \sqrt{3(1+2a^3)} ,\qquad 
    \frac{z_{\tau\tau}}{z} =\frac{3 (-1 + 154 a^3)}{4 \tilde{\eta}^2}.
\end{equation}
The modified Mukhanov-Sasaki equation (\ref{5:eq:v-modeeq}) now reads
% \begin{equation}
%     v_{k,\tau\tau}+\left(\frac{\tilde{\sigma}_S^4}{16 c^2 \tilde{\eta}^8} k^6+\frac{3}{4 \tilde{\eta}^2} - k^2-\frac{231}{4}c \tilde{\eta}^4\right)v_{k}=0,
%     \label{5:eq:v-modeeq-dust}
% \end{equation}
% \begin{equation}
%     v_{k,\tau\tau}+\left(\left(\frac{8\tilde{\sigma}_S}{H_l^{3}\tilde{\eta}^2}\right)^4 k^6+\frac{3}{4 \tilde{\eta}^2} - k^2-\frac{231}{16}\left(\frac{H_l^3 \tilde{\eta}^2}{8}\right)^2\right)v_{k}=0,
%     \label{5:eq:v-modeeq-dust}
% \end{equation}
\begin{equation}
    v_{k,\tau\tau}+\left(\left(\frac{\tilde{\sigma}_S}{H_l\, a}\right)^4 k^6+\frac{3}{4 \tilde{\eta}^2} - k^2-\frac{231}{16}\left(H_l\, a\right)^2\right)v_{k}=0,
    \label{5:eq:v-modeeq-dust}
\end{equation}
and we see that compared to (\ref{5:eq:v-modeeq-radiation}) there is an additional term coming from $z_{\tau\tau}/z$.

Assuming $\tilde{\sigma}_S\gg 1$, modes with
\begin{equation}
    k^2 \gg \mathcal{O}(10^{-1}) \frac{H_l^2}{\tilde{\sigma}_S^{4/3}}
    \label{5:eq:no-early-sh}
\end{equation}
exit the horizon at 
\begin{equation}
    \tilde{\eta}_k \sim \mathcal{O}(1)\frac{\tilde{\sigma}_S^{1/3}}{ H_l^{3/2}} \sqrt{k},
\end{equation}
and both the gradient term and the $1/\tilde{\eta}^2$ term never become important.

On the other hand, modes with wavelengths larger than (\ref{5:eq:no-early-sh}) already exit the horizon for a first time at 
\begin{equation}
    \tilde{\eta}_{1,k} \sim \mathcal{O}(1)\frac{\tilde{\sigma}_S^{2/3}}{H_l^2} k.
\end{equation}
Since $z_{\tau\tau}/z$ is changing sign at $\tilde{\eta}_0=(2^8/77)^{1/6}/H_l$, modes will at some point shortly re-enter the horizon due to the gradient term before finally re-exiting again. Expanding $z_{\tau\tau}/z$ around $\tilde{\eta}_0$, we can estimate the duration $\Delta \tilde{\eta}_k $ of the gradient instability region as
\begin{equation}
    k \Delta \tilde{\eta}_{k} \sim \mathcal{O}(1) \frac{k^3}{H_l^3} \ll \frac{\mathcal{O}(10^{-2}) }{\tilde{\sigma}_S^2}.
\end{equation}
During this short time span the mode function $v_k$ can only grow by a factor $\sim \exp\left(k \Delta \tilde{\eta}_{k} \right)$ which is completely negligible for $\tilde{\sigma}_S \gg \mathcal{O}(1)$.

Ignoring the effects of the gradient instability region, one would estimate the primordial spectrum of the longest wavelength modes $k \ll  H_l/\tilde{\sigma}_S^{2/3}$ which exit the horizon well before the end of inflation as
\begin{equation}
    \delta^2_{\zeta} \approx
  \mathcal{O}(1) \frac{k^3}{H_l}.
\end{equation}
Again, this is far being scale invariant with a spectral index $n_S-1 = 3$.
Since in this dust/MDM dominated case large wavelength modes go through several intermediate regions lasting only for a short time, such an analysis like in the radiation dominated case where the leading order solutions in different regions were continuously matched at the crossing between regions should be taken with caution. A full calculation of primordial spectra would require a numerical study beyond the scope of this paper.

%%%%%%%%%%%%%%%%%%%%%%%%%%%%%%%%%%%%%%%%%%%%%%%%%%%%%%%%%%%%%%%%%%%%%%
\section{Conclusions}
\label{5:sec:conclusions}
%%%%%%%%%%%%%%%%%%%%%%%%%%%%%%%%%%%%%%%%%%%%%%%%%%%%%%%%%%%%%%%%%%%%%%
The initial idea of ``Asymptotically Free Mimetic Gravity'' was to find a concrete modified theory of gravity with limiting curvature to address the singularity problem of GR. It has been successful in achieving this goal in a variety of settings, including both cosmological as well as black hole spacetimes. 
It was found that the concept of ``asymptotic freedom'', i.e. the vanishing of the $\Box\phi$ dependent gravitational constant at limiting curvature, becomes a crucial ingredient to resolve anisotropic singularities. 
Along the way it was realized that Ho\v{r}ava-like higher order spatial curvature terms can be added in a simple, covariant way to mimetic gravity with the goal of renormalizability.

Combining both ideas, in this paper I considered ``Asymptotically Free Mimetic Ho\v{r}ava Gravity''. In this theory, an initial stage of exponential expansion with graceful exit is a necessary feature of any non-singular modified flat Friedmann universe. It is a natural question to ask whether it could provide a full-blown inflationary scenario without inflaton. Since the existence of the inflationary background solution is independent of the matter content, as it is anyway suppressed by asymptotic freedom, there is no need to assume vanishing matter density during inflation. Thus, the idea of inflation driven by asymptotic freedom of gravity could open an interesting possibility to avoid the necessity of a reheating stage for the sake of populating the universe with matter after inflation.

In this work I analyzed stability of the inflationary background solutions under metric perturbations, considering only the degrees of freedom of pure mimetic gravity. It was found that a ghost instability is naturally avoided by any model with asymptotic freedom.
Although the gradient term of scalar perturbations is of the wrong sign throughout, its short era of domination is sandwiched between the domination of  higher order spatial curvature terms and the late time region where the scalar degree of freedom of mimetic gravity remains frozen forever. Thus, the gradient instability can be circumvented thanks to higher order spatial curvature terms. Under the condition that $\sigma_S\gg1/H_l$, I showed that the inflationary background solutions of asymptotically free models are free of any dangerous instability.

After passing stability tests, we have to ask if the primordial spectra produced by such an inflationary model can agree with CMB observations.
In this paper I showed that for the simplest one-component models the answer to this question is negative. The initially scale invariant spectra of both tensor and scalar perturbations can in general not be preserved until the horizon exit. For the concrete case of a radiation dominated inflationary background, the primordial spectra of the largest wavelength modes were found to be far from scale invariant with a large blue tilt and $n_T=n_S-1 =2$. Moreover, a small tensor-to-scalar ratio $r<\mathcal{O}(10^{-1})$ requires tuning of the higher order coefficients such that $\sigma_T\gg\mathcal{O}(10^{4})\sigma_S$.

However, it is clear that this analysis is incomplete, as any more realistic model would also have to include matter fluctuations. While the primordial spectrum of the dust-like scalar degree of freedom of mimetic gravity might be far from scale invariant, any other spectator field in the inflationary background solution would still acquire a nearly scale invariant spectrum. Depending on the details of the conversion process of perturbations between matter degrees of freedom and the conformal degree of freedom of mimetic gravity, one could thus speculate that in a ``curvaton''-like extension of the asymptotically free mimetic model, a nearly scale invariant primordial matter power spectrum would be obtainable. 
It remains an interesting open question whether it is possible to construct a viable inflationary scenario from shift-symmetric mimetic gravity.

\section*{Acknowledgements}
I would like to thank my supervisor Slava Mukhanov for his support and guidance throughout the years of my PhD. 

The author's work is supported by the Deutsche
Forschungsgemeinschaft (DFG, German Research Foundation) under Germany's
Excellence Strategy -- EXC-2111 -- 390814868.

%  \clearpage
\appendix
%%%%%%%%%%%%%%%%%%%%%%%%%%%%%%%%%%%%%%%%%%%%%%%%%%%%%%%%%%%%%%%%%%%%%%
\section{Calculation of second order actions}
\label{5:sec:appA}
%%%%%%%%%%%%%%%%%%%%%%%%%%%%%%%%%%%%%%%%%%%%%%%%%%%%%%%%%%%%%%%%%%%%%%
In this appendix I present the calculation of the second order actions used above. 
For generality, I will consider the action
\begin{equation}
\mathcal{S}_{g}=-\frac{1}{16\pi}\int\textup{d}^{4}x\sqrt{-g}\left\{L\left(X_{i}\right) +\lambda\left(  g^{\mu\nu}\phi_{,\mu}%
\phi_{,\nu}-1\right) \right\},   \label{5:eq:S-general}%
\end{equation}
where $L$ is a general function of the quantities 
\begin{equation}
    \begin{aligned}
X_{1} &= \Box \phi &\cong\quad& \kappa  \\
X_{2} &= \nabla^{\mu}\nabla^{\nu} \phi\nabla_{\mu}\nabla_{\nu}\phi -\tfrac{1}{3} (\Box \phi)^2 &\cong\quad& \tilde{\kappa}^{ij}\tilde{\kappa}_{ij} \\
X_{3} &= \widetilde{R}&\cong\quad& {^{3}\!R}  \\
    X_{4} &= -\widetilde{R}^{\mu\nu}\left(\nabla_{\mu}\nabla_{\nu}\phi- \tfrac{1}{3}\Box \phi\, g_{\mu\nu}\right)&\cong\quad&  {^{3}\!R}^{ij}\tilde{\kappa}_{ij}\\
    X_{5} &= \widetilde{R}^{\mu\nu}\widetilde{R}_{\mu\nu}-\tfrac{1}{3}\widetilde{R}^2&\cong\quad&  {^{3}\!\widetilde{R}^{ij}}\,{^{3}\!\widetilde{R}_{ij}} \\
    X_{6} &= -P^{\mu}_{\nu}\nabla_{\mu}\widetilde{R}\nabla^{\nu}\widetilde{R}&\cong\quad&\gamma^{kl}\bar{\nabla}_{k}({^{3}\!R})\bar{\nabla}_{l}({^{3}\!R})\\
    X_{7} &=-P^{\mu}_{\nu}P^{\alpha}_{\gamma}P^{\beta}_{\delta}\nabla_{\mu}\widetilde{R}^{\gamma\delta}\nabla^{\nu}\widetilde{R}_{\alpha\beta}+\tfrac{1}{3}X_{6} &\cong\quad&  \gamma^{kl}\bar{\nabla}_{k}({^3\!\widetilde{R}^{ij}})\,\bar{\nabla}_{l}({^3\!\widetilde{R}_{ij}})
\end{aligned}
\label{5:eq:Xi}
\end{equation}
where
\begin{align}
P_{\mu}^{\nu}&:=\delta_{\mu}^{\nu}-\phi_{,\mu}\phi^{,\nu} \\
\widetilde{R}_{\mu\nu}&:=P_{\mu}^{\alpha}P_{\nu}^{\beta}R_{\alpha\beta}+\nabla_{\alpha}\left(\phi^{,\alpha}\nabla_{\mu}\nabla_{\nu}\phi\right) \\
\widetilde{R}&:= 2\phi^{,\mu}\phi^{,\nu}G_{\mu\nu} - (\Box \phi)^2 + \nabla^{\mu}\nabla^{\nu}\phi \nabla_{\mu}\nabla_{\nu}\phi
\end{align}
The right column of (\ref{5:eq:Xi}), denoted by $\cong$, shows the quantities $X_i$ evaluated in comoving gauge $\phi =t$ where $N=1$, but the shift $N_i$ is still arbitrary. They are given in terms of the trace and trace-less part $\kappa = \gamma^{ij}\kappa_{ij}$, $\tilde{\kappa}_{ij} =  \kappa_{ij} - \tfrac{1}{3}\kappa \gamma_{ij}$, ${^3\!\widetilde{R}}= \gamma^{ij}\,{^3\!R_{ij}}$, ${^3\!\widetilde{R}_{ij}}={^3\!R_{ij}}- \tfrac{1}{3}{^3\!\widetilde{R}} \gamma_{ij}$  of the extrinsic curvature $\kappa_{ij} = \frac{1}{2N}\left(\dot{\gamma}_{ij}-\bar{\nabla}_{j}N_{i}-\bar{\nabla}_{i}N_{j}\right)$ and of the intrinsic Ricci tensor ${^3\!R_{ij}}$ of the spatial slices of (\ref{5:eq:ADM}), respectively. $\bar{\nabla}_i$ denotes the covariant derivative with respect to the spatial metric $\gamma_{ij}$. Indices on spatial tensors are raised with $\gamma^{ij}$.

Note that terms depending on the square of the spatial Cotton tensor $\widetilde{C}^{\mu}_{\nu}$ as introduced in (\ref{5:eq:Cotton}) 
are also covered by the general Lagrangian ansatz, since
\begin{equation}
   \widetilde{C}^{\mu}_{\nu}\widetilde{C}^{\nu}_{\mu}\cong {^{3}C}^{i}_{j}\,{^{3}C}^{j}_{i} \doteq X_7 + (\tfrac{1}{3}-\tfrac{3}{8})X_6+\tfrac{1}{2}X_3 X_5 +3\;{^{3\!}\widetilde{R}}^{i}_{j}{^{3}\!\widetilde{R}}^{j}_{k}{^{3}\!\widetilde{R}}^{k}_{i},
   \label{5:eq:Cottonexp}
\end{equation}
where $\doteq$ now denotes equality up to a total covariant spatial derivative $\bar{\nabla}$.
In an isotropic universe the trace-less part of $^{3}\!R_{ij}$ is first order in perturbations, hence the last term in (\ref{5:eq:Cottonexp}) is always higher order and does not contribute to the second order action.

Consider now perturbations around the general non-flat Friedmann background
\begin{equation}
    \mathrm{d}s^2 = dt^2-a^2(t) \eta_{ij}\mathrm{d}x^i \mathrm{d}x^j, \qquad \eta_{ij}=\frac{\delta_{ij}}{\left(1+\tfrac{\varkappa}{4}(x^2+y^2+z^2)\right)^2}
\end{equation}
in comoving gauge $\phi =t$. The metric perturbations of (\ref{5:eq:ADM}) are then further decomposed as
\begin{equation}
    \gamma_{ij}=a^2(t)\left(e^{-2\psi}{\eta_{ij}} - 2D_iD_jE -2D_{(i}F_{j)} + (e^{h})_{ij}\right),\qquad N_{i} = \chi_{,i}-a S_{i}
\end{equation}
where
\begin{equation}
   \eta^{ij}D_jF_{i} = 0, \qquad \eta^{ij}h_{ij} = 0, \qquad \eta^{jk}D_{k}h_{ij} = 0 , \qquad \eta^{ij}D_{j}S_{i}=0
\end{equation}
and $D_{i}$ denotes the covariant derivative associated with $\eta_{ij}$.

\paragraph{Background.} Variation of the zeroth order action ${^{(0)}\mathcal{S}} = {^{(0)}\mathcal{S}_{g}+ {^{(0)}\mathcal{S}_{m}}}$ with respect to $a^3$ yields
\begin{equation}
\delta_{a^3}{^{(0)}\mathcal{S}}  = -\frac{1}{16\pi}\int\!\mathrm{d}^4x\, \sqrt{\eta}\left[  L -\tfrac{1}{a^3}\partial_{0}\left(a^3 L_{1}\right) -4L_{3}\tfrac{\varkappa}{a^2}-16\pi   p \right]\, \delta( a^3)
\end{equation}
where subscripts $L_{i}$ denote derivatives of $L$ with respect to $X_i$ evaluated on the background. Here it was used that for homogeneous, isotropic matter 
\begin{equation}
    \delta \mathcal{S}_{m} = \tfrac{1}{2}\int\!\mathrm{d}^4x \sqrt{-g}\, T_{\mu\nu}\delta g^{\mu\nu} = \int\!\mathrm{d}^4x \sqrt{\eta}\, p\, \delta( a^3).
\end{equation}
The background equation of motion is hence
\begin{equation}
     L -\frac{1}{a^3}\partial_{0}\left(a^3 L_{1}\right) -4L_{3}\frac{\varkappa}{a^2}= 16\pi  p .
     \label{5:eq:backgroundeom}
\end{equation}
Using the continuity equation $\frac{1}{a^3 }\partial_{0}\left(a^3 \varepsilon\right)= - 3\frac{\dot{a}}{a} p$,
the first integral of (\ref{5:eq:backgroundeom}) becomes the modified Friedmann equation
\begin{equation}
    \frac{1}{2}\left(\kappa L_{1}-L \right)=\frac{c_{\textup{MDM}}}{u}+8\pi  \varepsilon = 8 \pi \varepsilon,
\end{equation}
where the constant of integration $c_{\textup{MDM}}$ describes the contribution of mimetic matter.

\paragraph{Tensor perturbations.}
Tensor perturbations are parametrized by
\begin{equation}
    \gamma_{ij} = a^2\left(\eta_{ij}+h_{ij}+\frac{1}{2}h_{i}^{k}h_{kj}\right), \qquad N_{i} = 0,
\end{equation}
where indices on $h_{ij}$ are raised with $\eta^{ij}$. With this parametrization the inverse spatial metric is
\begin{equation}
    \gamma^{ij} = \frac{1}{a^2}\left(\eta^{ij}-h^{ij}+\frac{1}{2}h^{ik}h_{k}^{j}\right) +\mathcal{O}(h^3)
\end{equation}
and it holds that $\sqrt{\gamma}= a^3\sqrt{\eta}$ and $\kappa = 3\frac{\dot{a}}{a}$ are still homogeneous up to $\mathcal{O}(h^3)$. The extrinsic curvature and intrinsic Ricci scalar up to second order are
\begin{align}
    \kappa_{ij} &= \frac{\dot{a}}{a}\gamma_{ij}+\frac{1}{2}a^2\left(\dot{h}_{ij}+\frac{1}{2}\dot{h}_{ik}h^{k}_{~j}+\frac{1}{2}h_{ik}\dot{h}^{k}_{~j}\right),
\\
    {^3\!R} &\doteq\frac{1}{a^2}\left( 6\varkappa- \frac{1}{4}D_{k}h^{ij} D^{k}h_{ij}-\frac{1}{2}\varkappa\, h^{ij}h_{ij} \right),
\end{align}
where $\doteq$ now denotes equality up to a total covariant spatial derivative $D$. The spatial Ricci tensor and its trace-less part up to first order are given by
\begin{equation}
  {^3\!R}_{ij} = 2\varkappa \eta_{ij}- \frac{1}{2}\Delta h_{ij}+3\varkappa h_{ij}, \qquad 
  {^3\!\widetilde{R}}_{ij} = - \frac{1}{2}\Delta h_{ij}+\varkappa h_{ij}, 
\end{equation}
where $\Delta = \eta^{ij} D_i D_j$.
The quantities (\ref{5:eq:Xi}) expanded to second order are
\begin{equation}
    \begin{aligned}
X_{1} &= 3 \tfrac{\dot{a}}{a} \\
X_{2} &=\tfrac{1}{4} \dot{h}^{ij}\dot{h}_{ij} \\
X_{3} &\doteq \tfrac{1}{a^2}\left( 6\varkappa- \tfrac{1}{4}D_{k}h^{ij} D^{k}h_{ij}-\tfrac{1}{2}\varkappa\, h^{ij}h_{ij} \right) \\
    X_{4} &\doteq \tfrac{1}{8a^2} \partial_{0}\left( D_{k}h^{ij}D^{k} h_{ij} +2\varkappa h^{ij}h_{ij}\right) \\
    X_{5} &\doteq \tfrac{1}{a^4}\left(\tfrac{1}{4}\Delta h^{ij}\Delta h_{ij}+ \varkappa D_{k}h^{ij} D^{k}h_{ij}+\varkappa^2 h^{ij}h_{ij}\right) \\
    X_{6} &= 0 \\
    X_{7} &\doteq \tfrac{1}{a^6}\left(\tfrac{1}{4}D_{k}\Delta h^{ij} D^{k}\Delta h_{ij}+ \varkappa \Delta h^{ij}\Delta h_{ij}+\varkappa^2 D_{k}h^{ij} D^{k}h_{ij}\right)
\end{aligned}
\end{equation}
and we find the second order action
\begin{align}
 {^{(2)}\mathcal{S}}_{h} = -\frac{1}{64\pi}
\int\!\mathrm{d}^4x &\sqrt{\eta}\, a^3  \bigg \{ L_{2}\,\dot{h}^{ij}\dot{h}_{ij} - \left(L_{3}+\tfrac{1}{2a}\partial_{0}\left(aL_{4}\right)-\tfrac{2\varkappa}{a^2}L_{5}\right)\tfrac{2\varkappa}{a^2}\,h^{ij}h_{ij}  + \\
& + \left(L_{3}+\tfrac{1}{2a}\partial_{0}\left(aL_{4}\right)-\tfrac{4\varkappa }{a^2}\left(L_{5}+\tfrac{\varkappa}{a^2}L_7\right)\right)\tfrac{1}{a^2}h^{ij} \Delta h_{ij}+ \phantom{\bigg \} } 
\nonumber\\ &
+\frac{L_{5}+\tfrac{4\varkappa}{a^2}L_{7}}{a^4}h^{ij}\Delta^2 h_{ij}-\frac{L_{7}}{a^6}h^{ij}\Delta^3 h_{ij}\bigg \}. \nonumber
\end{align}
Subscripts $L_{i}$ denote derivatives of $L$ with respect to $X_i$ evaluated on the background.
\paragraph{Scalar perturbations.}
The scalar metric perturbations in comoving gauge are parametrized through $\zeta$ and $\chi$ by
\begin{equation}
    \gamma_{ij}=a^2(t)e^{2\tilde{\zeta}}\delta_{ij},\qquad\qquad  N_{i} = \chi_{,i},
\end{equation}
where
\begin{equation}
    \tilde{\zeta} = \zeta-\ln\left(1+\tfrac{\varkappa}{4}(x^2+y^2+z^2)\right).
\end{equation}
The metric determinant is given by $\sqrt{\gamma} = a^3 e^{3\tilde{\zeta}}=a^3\sqrt{\eta}\, e^{3\zeta}$
and the trace of extrinsic curvature is
\begin{equation}
    \kappa = 3\frac{\dot{a}}{a}+3\dot{\zeta}-\bar{\Delta}\chi,
\end{equation}
where the Laplacian $\bar{\Delta}=\gamma^{ij}\bar{\nabla}_{i}\bar{\nabla}_{j}$ of the covariant derivative $\bar{\nabla}$ with respect to $\gamma_{ij}$ is given by
\begin{equation}
    \bar{\Delta} \chi = \tfrac{1}{a^2} \Delta\chi-\tfrac{2}{a^2}\zeta \Delta\chi+\tfrac{1}{a^2} \eta^{ij}\zeta_{,i}\chi_{,j}+\mathcal{O}\left(\chi \zeta^2\right).
\end{equation}
To first order, the traceless part of $\kappa_{ij}$ is given entirely through $\chi$ as
\begin{equation}
    \tilde{\kappa}_{ij} = -\bar{\nabla}_{j}\bar{\nabla}_{i} \chi + \tfrac{1}{3}\bar{\Delta} \chi \gamma_{ij} =-D_{j}D_{i} \chi + \tfrac{1}{3}\Delta \chi \eta_{ij} +\mathcal{O}(\zeta\chi).
\end{equation}
The spatial metric $\gamma_{ij}$ is conformally flat which simplifies the calculation of 
\begin{align}
    {^3\!R_{ij}} &= -\left(\tilde{\zeta}_{,ij}- \tilde{\zeta}_{,i}\tilde{\zeta}_{,j}\right)-\left(\tilde{\zeta}^{,k}_{~k}+ \tilde{\zeta}^{,k}\tilde{\zeta}_{,k}\right)\gamma_{ij}\\&= 2\varkappa \eta_{ij}-\left(D_jD_i\zeta+\Delta \zeta\eta_{ij}\right)+\left( \zeta_{,i}\zeta_{,j}-
     \eta^{kl}\zeta_{,k}\zeta_{,l}\,\eta_{ij}\right) \nonumber
\\
    {^3\!R} &= \frac{e^{-2\zeta}}{a^2}\left(6\varkappa-4\Delta\zeta-2\eta^{ij}\zeta_{,i}\zeta_{,j}\right)\\
           &=\frac{1}{a^2}\left(6\varkappa-4 \Delta\zeta-12\varkappa \zeta +12\varkappa \zeta^2 +8\zeta\Delta\zeta-2\eta^{ij}\zeta_{,i}\zeta_{,j}\right) +\mathcal{O}\left(\zeta^3\right)\nonumber\\
               {^3\!\widetilde{R}_{ij}} &=-\left(D_j D_i\zeta-\tfrac{1}{3}\Delta \zeta\eta_{ij}\right)+\mathcal{O}(\zeta^2)
\end{align}
where $\Delta = \eta^{ij}D_{i}D_{j}$. Combining these results, the quantities (\ref{5:eq:Xi}) expanded to second order read
\begin{equation}
    \begin{aligned}
X_{1} &= 3 \tfrac{\dot{a}}{a} +3\dot{\zeta} -\tfrac{1}{a^2}\Delta\chi+\tfrac{2}{a^2}\zeta\Delta\chi -\tfrac{1}{a^2} \left(D\zeta\right)^2\\
X_{2} &\doteq  \tfrac{1}{ a^4}\left(\tfrac{2}{3}(\Delta \chi )^2 - 2\varkappa\left(D\chi\right)^2\right)\\
X_{3} &= \tfrac{1}{a^2}\left(6\varkappa-4 \Delta\zeta-12\varkappa \zeta +12\varkappa \zeta^2 +8\zeta\Delta\zeta-2\left(D\zeta\right)^2\right)\\
    X_{4} &\doteq \tfrac{1}{a^4}\Delta \chi \left(\tfrac{2}{3}\Delta\zeta + 2\varkappa \zeta\right) \\
    X_{5} &\doteq \tfrac{1}{a^4}\left(\tfrac{2}{3}(\Delta \zeta)^2-2\varkappa \left(D\zeta\right)^2\right)\\
    X_{6} &\doteq \tfrac{16}{a^6}\left(\left(D \Delta\zeta\right)^2-6\varkappa\left(\Delta\zeta\right)^2+9\varkappa^2\left(D\zeta\right)^2\right) \\
    X_{7} &\doteq \tfrac{1}{a^6}\left(\tfrac{2}{3}\left(D \Delta\zeta\right)^2-6\varkappa\left(\Delta\zeta\right)^2+12 \varkappa^2 (D\zeta)^2\right) 
\end{aligned}
\end{equation}
with the notation $\left(D\zeta\right)^2 \equiv \eta^{ij}\zeta_{,i}\zeta_{,j}$.

Neglecting matter fluctuations of $\mathcal{S}_{m} = \int\!\mathrm{d}^4x \sqrt{\eta}\,a^3\,e^{3\zeta}  p$ and using the background equation of motion (\ref{5:eq:backgroundeom}), the total second order action ${^{(2)}\mathcal{S}} = {^{(2)}\mathcal{S}_{g}+ {^{(2)}\mathcal{S}_{m}}}$ becomes 
\begin{align}
{^{(2)}\mathcal{S}}& = -\frac{1}{32 \pi}
\int\! \mathrm{d}^4x\sqrt{\eta} \, a^{3}\bigg{\{}
    -\left[L_3-\tfrac{3}{a}\partial_{0}\left(a L_{13} \right) -\tfrac{12\varkappa}{a^2} L_{33} \right]\, \tfrac{12\varkappa}{a^2} \zeta^2 +\\
   &+ L_{11}9\dot{\zeta}^2-\left[L_{3} -  \tfrac{3}{a} \partial_{0}\left(a L_{13}\right) - \tfrac{\varkappa}{a^2}(24L_{33}+L_{5})+\tfrac{6\varkappa^2}{a^4}(12L_{6}+L_7) \right]  \tfrac{4}{a^2}\,\zeta \Delta \zeta +\phantom{\bigg{\}}}\nonumber\\
   &+\left[16L_{33}+\tfrac{4}{3}L_{5}-\tfrac{12\varkappa}{a^2}\left(16L_{6}+L_{7}\right)\right]\tfrac{1}{a^4}\,\zeta \Delta^2 \zeta -\left[32L_{6}+\tfrac{4}{3}L_{7}\right]\tfrac{1}{a^6}\,\zeta \Delta^3 \zeta \phantom{\bigg{\}}}\nonumber\\
   &+ \left[(8L_{13}+\tfrac{4}{3}L_{4})\left(3\varkappa\zeta+ \Delta\zeta\right)-6L_{11}a^2 \dot{\zeta}+ (L_{11}+\tfrac{4}{3}L_{2})\Delta \chi+ 4\varkappa L_{2} \chi \right]\tfrac{1}{a^4}\Delta\chi
     \bigg{\}}. \nonumber
\end{align}
Subscripts $L_{i}$ denote derivatives of $L$ with respect to $X_i$ evaluated on the background.

Variation with respect to $\Delta \chi$ yields 
\begin{equation}
    (L_{11}+\tfrac{4}{3}L_{2})\Delta \chi+ 4\varkappa L_{2} \chi = 3L_{11}a^2 \dot{\zeta}-(4L_{13}+\tfrac{2}{3}L_{4})\left(3\varkappa\zeta+ \Delta\zeta\right).
\end{equation}

In the spatially flat case the second order action for $\zeta$ after integrating out $\chi$, assuming $L_{11}+\tfrac{4}{3}L_{2}\neq0$, becomes
\begin{align}
 {^{(2)}\mathcal{S}_{\zeta}}=-&\frac{1}{32\pi} \int\!\mathrm{d}^4x \, a^{3}\bigg{\{}
    \frac{12 L_{11}L_{2}}{L_{11}+\tfrac{4}{3}L_{2}}\,\dot{\zeta}^2-\left[L_{3} -  \tfrac{1}{a} \partial_{0}\left(a\frac{4L_{2}L_{13}-\tfrac{1}{2}L_{11}L_{4}}{L_{11}+\tfrac{4}{3}L_{2}}\right)\right]  \tfrac{4}{a^2}\,\zeta \Delta \zeta\phantom{\bigg{\}}}\nonumber\\
   &+\left[16L_{33}+\tfrac{4}{3}L_{5}- \frac{(4L_{13}+\tfrac{2}{3}L_{4})^2}{L_{11}+\tfrac{4}{3}L_{2}}\right]\tfrac{1}{a^4}\, \zeta \Delta^2\zeta -\left[32L_{6}+\tfrac{4}{3}L_{7}\right]\tfrac{1}{a^6}\,\zeta \Delta^3 \zeta \bigg{\}} .
\end{align}
\paragraph{NB.} 
Note that in the case $L_{33}=L_{5}=L_{6}=L_{7}=0$ it is tempting to make the choice, $L_{4} = -6L_{13}$ such that
\begin{align}
 {^{(2)}\mathcal{S}_{\zeta}}=-\frac{1}{32\pi} \int\!\mathrm{d}^4x\, a^{3}\bigg{\{}
   & \frac{12 L_{11}L_{2}}{L_{11}+\tfrac{4}{3}L_{2}}\,\dot{\zeta}^2-\left[L_{3} - 3 \tfrac{1}{a} \partial_{0}\left(a L_{13}\right)\right]  \tfrac{4}{a^2}\,\zeta \Delta \zeta\bigg{\}} .
\end{align}
However, note that in this case a gradient instability can only be prevented at the expense of introducing a gradient instability for tensor perturbations.

%%%%%%%%%%%%%%%%%%%%%%%%%%%%%%%%%%%%%%%%%%%%%%%%%%%%%%%%%%%%%%%%%
\section{Stability analysis of higher order spatial curvature bounces}
\label{5:sec:appB} 
%%%%%%%%%%%%%%%%%%%%%%%%%%%%%%%%%%%%%%%%%%%%%%%%%%%%%%%%%%%%%%%%%
As seen in \cite{Chamseddine:2019fog}, the goal of singularity resolution in spatially non-flat universes requires the introduction of a scalar spatial curvature dependent potential. By isotropy of the background, there is a degeneracy of higher order spatial curvature terms that will lead to the same background dynamics. In the following I will consider the generalization
\begin{equation}
    \mathcal{L} = \mathcal{L}_{\textup{nhd}}+\mathcal{L}_{\textup{hd}}+ V(\widetilde{R}) + \alpha^4\, \widetilde{R}\left( \widetilde{R}^{\mu\nu}\widetilde{R}_{\mu\nu}-\tfrac{1}{3}\widetilde{R}^2\right),
\end{equation}
where now $\sigma_T$, $\sigma_S$ and $\alpha$ can depend on $\widetilde{R}$. Note that in the spatially flat case this Lagrangian is equivalent to (\ref{5:eq:S}).

\paragraph{Background.} The modified Friedmann equation with non-vanishing spatial curvature becomes
\begin{equation}
 \tfrac{2}{3}\left(\kappa\ell'-\ell\right) =\tfrac{1}{2}\left(V\left({^{3}\!R}\right)-{^{3}\!R}\right) +8\pi \varepsilon.
 \label{5:eq:mFriedmann-nonflat}
\end{equation}
It is independent of the higher order terms $\sigma_T$, $\sigma_S$ and $\alpha$. This modified Friedmann equation has been discussed in \cite{Chamseddine:2019fog}.
A bounce is made possible if the term $V\left({^{3}\!R}\right)$ is negative and dominating over $\varepsilon$ at small scale factor. Assuming a cubic potential,
\begin{equation}
    V\left({^{3}\!R}\right) =- 6\,\delta^4\left( \frac{\varkappa}{6}{^3\!R}\right)^{3} = - \frac{6\,\delta^4}{a^6},
    \label{5:eq:Vcubic}
\end{equation}
this will be satisfied, provided that the matter equation of state satisfies $w<1$. 
The bounce then happens at the minimal value of the scale factor 
\begin{equation}
    a_{\min} = \left(\frac{1+3w}{2}\right)^{\tfrac{1+w}{1-w}} \left(\frac{\sqrt{c}}{H_{l}}\right)^{-\tfrac{1+3w}{3(1-w)}} \delta^{\tfrac{4}{3(1-w)}},
\end{equation}
where (\ref{5:eq:etot}) was assumed. In the case $\varkappa =1$ a re-collapse happens at
\begin{equation}
    a_{\max} =\left(\frac{1+3w}{2}\right)^{-\tfrac{3(1+w)}{1+3w}} \frac{\sqrt{c}}{H_l}.
\end{equation}
\paragraph{Second order actions.} The second order action for tensor perturbations is
\begin{align}
{^{(2)}\mathcal{S}}_{h} = \frac{1}{64\pi}\int\!\mathrm{d}^4x\sqrt{\eta}\, a^3  \bigg \{ &f\,\dot{h}^{ab}\dot{h}_{ab} - \left(1-V'+\tfrac{6\varkappa^2}{a^4}(\sigma_T^4+2\alpha^4)\right)\tfrac{2\varkappa}{a^2}\,h^{ab}h_{ab}   \label{5:eq:2Sh-nonflat}\\
& - \left(1-V'+\tfrac{8\varkappa^2}{a^4}(2\sigma_T^4+3\alpha^4)\right)\tfrac{1}{a^2}(D h_{ab})^2 
 \nonumber\\ &
-(7\sigma_T^4+6\alpha^4)\tfrac{\varkappa}{a^6}(D^2 h_{ab})^2-\frac{\sigma_T^4}{a^6}(D^3h_{ab})^2\bigg \}, \nonumber
\end{align}
where $(Dh_{ab})^2 \equiv \eta^{cd}D_{c}h^{ab}D_{d}h_{ab}$, $(D^2h_{ab})^2 \equiv \Delta h^{ab}\Delta h_{ab}$, $(D^3h_{ab})^2 \equiv \eta^{cd}D_{c}\Delta h^{ab}D_{d}\Delta h_{ab}$.

Neglecting matter perturbations, the second order action for scalar perturbations, after integrating out $\chi$ is
\begin{align}
{^{(2)}\mathcal{S}_{\zeta}} = \frac{1}{8\pi}
\int\!\mathrm{d}^4x\sqrt{\eta} \, a^{3}\bigg{\{}
   &\dot{\zeta}\,\frac{3\ell'' f \left(\Delta+3\varkappa\right)}{(\ell''-f)\Delta-3f\varkappa}\,\dot{\zeta}- \left[\tfrac{1}{3}(1-V')+4V''\tfrac{\varkappa}{a^2} \right] \tfrac{\varkappa}{a^2}\, 9 \zeta^2 \label{5:eq:2Szeta-nonflat} \\
   &+\left[1-V'  + \tfrac{3\varkappa}{a^2}(8V''+\tfrac{\varkappa}{a^2}(2\alpha^4-3\sigma_S^4) \right] \tfrac{1}{a^2}\,(D\zeta)^2 +\phantom{\bigg{\}}}\nonumber\\
   &-\left[4V''+\tfrac{2\varkappa}{a^2}(\alpha^4-3\sigma_S^4)\right]\tfrac{1}{a^4}\, (D^2\zeta)^2 -\tfrac{\sigma_S^4}{a^6}\,(D^3 \zeta)^2  \bigg{\}}. \nonumber 
\end{align}

\paragraph{Ghost instability?}
The condition for no ghost instability of the mode characterized by the eigenvalue $-k^2$ of the curved Laplacian $\Delta$ can be written as
\begin{equation}
    \frac{\ell'' f \left(k^2-3\varkappa\right)}{(\ell''-f)k^2+3f\varkappa} >0.
\end{equation}
In the case $\ell''\neq f$, the condition for short wavelength modes $k^2 \gg \varkappa$ is the same as (\ref{5:eq:ghostfree1}) in the spatially flat case. 
On the other hand, in the limit $\ell''-f\to0$, $f\to 1$ (which applies also in the region around the bounce where $H$ vanishes) the condition becomes
\begin{equation}
    \varkappa k^2 > 3.
\end{equation}
In an open universe $\varkappa= -1$ the spectrum of the curved Laplacian is continuous and bounded by $k^2\geq1$. Hence all modes suffer from a ghost instability.
In a closed universe $\varkappa= 1$, however, the eigenvalues of $\Delta$ are discrete and given by \cite{Akama:2018cqv}, \cite{Abbott:1986ct}
\begin{equation}
    k^2 = n^2 -\varkappa, \qquad n\in \mathbb{N}_{\geq1}.
\end{equation}
The discreteness of the spectrum is due to the periodic boundary conditions that eigenfunctions have to satisfy. 
Note that only the modes $n=1,2$ would suffer from a ghost instability.
However, these two longest wavelength modes can be shown to correspond to pure gauge terms \cite{Lifshitz:1963ps}, \cite{Bardeen:1980kt}.
In conclusion, in the closed case $\varkappa =1$ there is no ghost instability.

\paragraph{Gradient instability?}
Let us now consider $\varkappa=1$ and take for simplicity the cubic potential (\ref{5:eq:Vcubic}) and constant $\sigma_S^4\geq0$, $\sigma_T^4\geq0$, $\alpha^4\geq0$.
In this case tensor perturbations do not exhibit any instability and the sixth order term in (\ref{5:eq:2Szeta-nonflat}) does not lead to an instability of scalar perturbations. The condition following from the right sign of the forth order term reads
\begin{equation}
\alpha^4 \geq 3\sigma_S^4+2\delta^4.
\label{5:eq:alphacondition}
\end{equation}
From the gradient term we can read off that there is a gradient instability wherever 
\begin{equation}
1+ \tfrac{3}{a^4}(2\alpha^4-3\sigma_S^4-7\delta^4)  >0.
\end{equation}
Using (\ref{5:eq:alphacondition}), we see that if $\delta^4\leq2\sigma_S^4$ this condition is satisfied at all times and there is a gradient instability throughout.
Conversely, if $\delta^4>2\sigma_S^4$ it is possible to avoid the gradient instability in the region around the bounce, provided that
\begin{equation}
    a_{\min}^4 < 3\left(7\delta^4+3\sigma_S^4-2\alpha^4\right).
\end{equation}
In order to avoid the gradient instability at all times, it would be necessary to have
\begin{equation}
    a_{\max}^4 < 3\left(7\delta^4+3\sigma_S^4-2\alpha^4\right) \overset{!}{<} 9\left(\delta^4-2\sigma_S^4\right),
\end{equation}
where for the last inequality (\ref{5:eq:alphacondition}) was used.
Note, however, that in this case
\begin{equation}
    \left(\frac{a_{\max}}{a_{\min}}\right) \sim \left(\frac{a_{\max}}{\delta}\right)^{\tfrac{4}{3(1+w)}} \overset{!}{<} \mathcal{O}(10),
\end{equation}
and hence a gradient instability region is unavoidable for any universe that undergoes any significant amount of expansion.

\end{document}